\documentclass{article}
\usepackage[natbib=true,style=numeric, sorting=none]{biblatex}
\usepackage{amssymb}
\usepackage{amsmath}
\usepackage{graphicx} 
\usepackage{enumitem}
\usepackage{braket}
\usepackage{color}
\newcommand{\TotalStructs}{150\phantom{ } }
\newcommand{\FlatStructs}{48\phantom{ } }

\begin{document}
\title{First Principles Study of Electronic Structure and Transport in Graphene Grain Boundaries}
%\title[Transport and Localized States in Graphene Grain Boundaries: First Principles Calculations} 
%Electronic Structure, Transport and Spin-polarization in Graphene Grain Boundaries]{Electronic Structure, Transport and Spin-polarization in Graphene Grain boundaries}
\author{Aleksander Bach Lorentzen$^{1}$, Fei Gao$^2$, Peter Bøggild$^{1}$, \\Antti-Pekka Jauho$^1$ and Mads Brandbyge$^{1}$}

$^1$ Department of Physics, Technical University of Denmark, 2800 Kongens Lyngby, Denmark\\
$^2$ Donostia International Physics Center (DIPC), 20018 Donostia-San Sebasti\'an, Spain
%abalo@dtu.dk
\begin{abstract}
%Old Abstract
%The local electronic structure of graphene grainboundaries (GBs) are investigated together with their elastic conductive properties in a screening study of many different GBs that has been generated from first-principles DFT relaxation. Intricate electronic structure is present around all GBs in our dataset and all exhibit a quasi 1D bandstructure transverse to the GB which could be sensitive to gating, while others have non-dispersive states as well. These states are characterized in a quasi-particle scheme. Four different types of GBs appear from the dataset when considering their conductive properties: Transparent, opaque, insulating and spin-filtering GBs. Availability of states to tunnel from and to in conjunction with interference effects and out-of-plane buckling produce these different classes. STM measurements are furthermore simulated on these various types of GBs.
% PB abstract
%A large number of 
Grain boundaries play a major role for electron transport in graphene sheets grown by chemical vapor deposition. Here we investigate the electronic structure and transport properties of idealized graphene grain boundaries (GBs) in bi-crystals using first principles density functional theory (DFT) and non-equilibrium Greens functions (NEGF).
We generated \TotalStructs different grain boundaries using an automated workflow where their geometry is relaxed with DFT. We find that the GBs generally show a quasi-1D bandstructure along the GB.
We group the GBs in four classes based on their conductive properties: transparent, opaque, insulating, and spin-polarizing and show how this is related to angular mismatch, quantum mechanical interference, and out-of-plane buckling. Especially, we find that spin-polarization in the GB correlates with out-of-plane buckling. We further investigate the characteristics of these classes in simulated scanning tunnelling spectroscopy and diffusive transport along the GB which demonstrate how current can be guided along the GB.  
%arise from tunneling states, interference effects, and out-of-plane buckling.
\end{abstract}
%\keyword{Graphene, Grainboundaries, defect bands, gate tuneable, spin-polarising, STM simulation, Quasiparticle states, workflow, non-equilibrium.}

%\submitto{\TDM}
\maketitle
%\ioptwocol
\section{Introduction}
% Old
%Grain boundaries (GBs) are linedefects that are abundant in graphene grown by chemical vapor deposition (CVD) on common catalytic surfaces such as Cu because of the way the graphene grows from individual nucleation centres.\cite{biro2013grain, tapaszto2012mapping} 

\begin{figure}[h]
    \centering
    \includegraphics[width=0.7\textwidth]{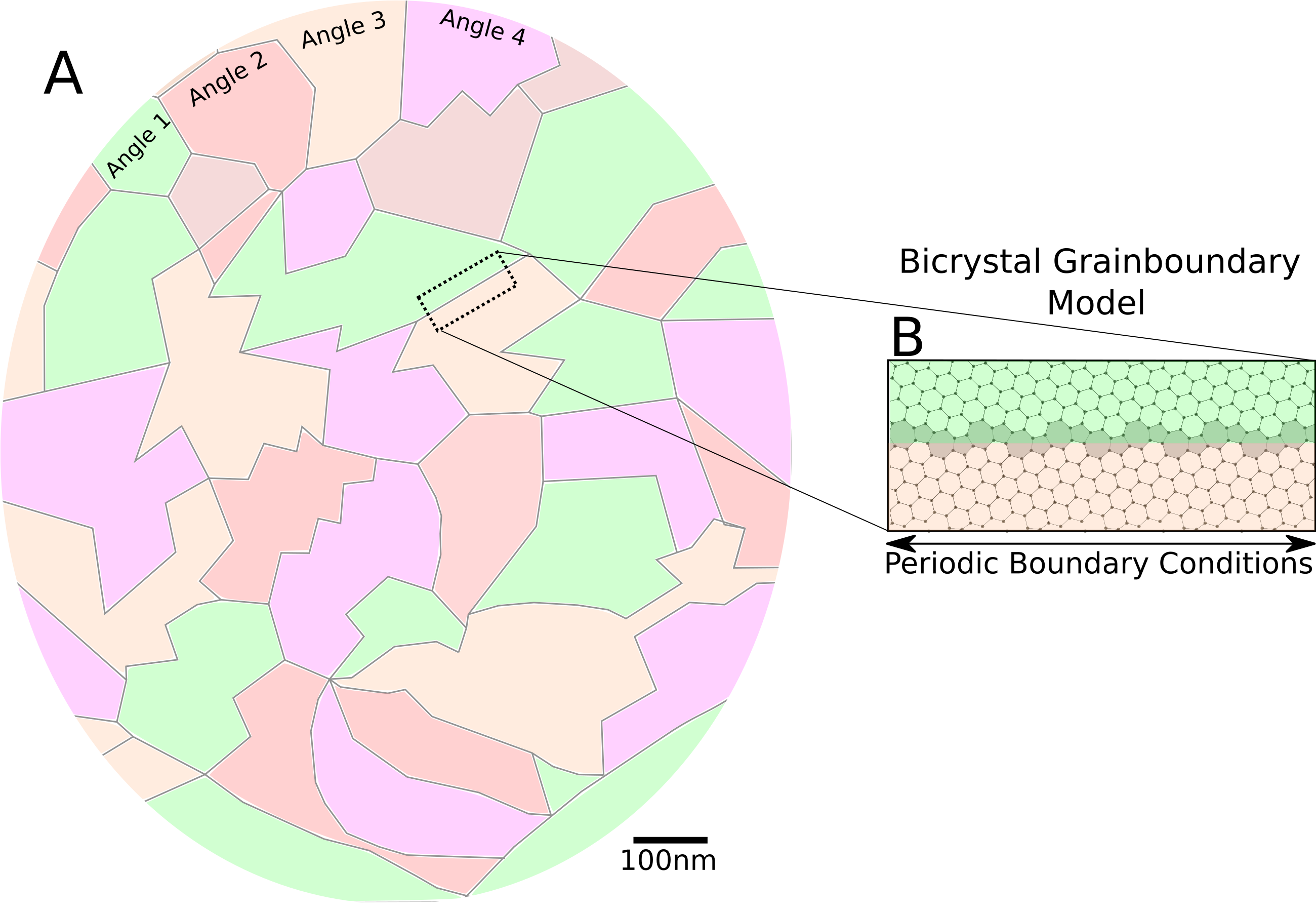}
    \caption{Cartoon illustration of grain formation in CVD-grown graphene,  inspired by Ref. \cite{huang2011grains}. The four different colors indicate different angles of orientation of the graphene lattice in each grain. Part B) is not to scale in A)}
    \label{fig:GrainIllust}
\end{figure}

% PB version
Grain boundaries (GBs) are line defects that occur in monolayer graphene grown by chemical vapor deposition (CVD) on common catalytic surfaces such as Cu where non-epitaxial growth from multiple nucleation centers causes orientation mismatch at the boundary between individual grains or domains\cite{biro2013grain, tapaszto2012mapping, huang2011grains}. This is illustrated in Fig. \ref{fig:GrainIllust}.
In these one-dimensional regions where the grains interface, the two graphene lattices deform and reconstruct into disordered structures which exhibit varying degrees of order along the boundary\cite{biro2013grain, ophus2015large, yang2014periodic, tison2014grain, kim2011grain, huang2011grains, paez2015electronic}.
%Old
%As CVD growth on a suitable substrate is one of the ways to produce large quantities of graphene for possible use in electronics.\cite{zhang2013review, de2009synthesis} This means the way the GB's affects the electronic transport properties of graphene are important to understand. Previously geometrical effects of how two lattices that are twisted relative to each other have been considered.\cite{yazyev2010electronic} On the other hand, the atomic structure making up the grain-boundary also has a significant effect on the transport properties, as interference between scattered waves can facilitate a strongly reflecting grain-boundary. Furthermore, grain-boundaries can facilitate a localized voltage drop if a bias is applied.\cite{clark2013spatially} This means non-equilibrium effects can be expected at a GB.
% PB version
As CVD growth of polycrystalline graphene is the most common means of producing large quantities of graphene for use in electronics or optoelectronics\cite{zhang2013review, de2009synthesis, huang2022graphene}, the impact of GBs on the overall and local transport properties has been an area of research for more than a decade\cite{xu2018enhancing,ihnatsenka2013electron, yasaei2014chemical, li2014grain, malola2010structural, mesaros2010electronic, ma2014evidence, cockayne2011grain, zhang2012electronic, hus2017spatially, tsen2012tailoring, cummings2014quantum,cummings2014charge}. In addition, the one-dimensional interface between two misoriented graphene grains has been shown to have several surprising and useful properties, including Yu-Shiba-Rusinov states and spontaneous time-reversal symmetry breaking\cite{hsieh2021spontaneous, cortes2021observation}, localized voltage drops\cite{clark2013spatially, bevan2014first}, optical response\cite{fei2013electronic, duong2012probing} and they could be used as electron wave guides\cite{mark2012forming}.
Previous studies have considered the geometric effects in two lattices twisted relative to each other, which gives rise to two different types of GBs; conducting and non-conducting, depending on if the GB is momentum mismatched or not\cite{yazyev2010electronic}. On the other hand, quantum interference effects related to how the carbon atoms bond and to the local electronic structure at the GB in general can also have significant influence on the conduction of electrons\cite{solomon2008understanding, cardamone2006controlling}.
%On the other hand, the atomic structure of the grain-boundary also has a significant effect on the transport properties, as interference between scattered waves can facilitate a strongly reflecting grain-boundary. 
%Furthermore, GBs can facilitate a localized voltage drop if a bias is applied.\cite{clark2013spatially} This means non-equilibrium effects can be expected at a GB.

There are many different possible ways in which two graphene sheets with a different orientation can interface and reconstruct, and the details of how the atomic structure influences the sheet conductivity of polycrystalline graphene is still not well understood. Here we take an approach inspired by "big data" and generate a set of \TotalStructs GBs to get a deeper insight into the effects of the atomic arrangements, their classifications and trends.

%In this paper it is furthermore found that the GB have an a further intricacy in its electronic structure because of dispersive transversal bands of localised/defect states, which depends on the atomic structure at the grainboundary. The literature on the 1D systems of electrons is stretches many years back in time, being a possible host for Luttinger-liquids, ...??

\section{Methods}
\subsection{From pristine graphene to a grain boundary}
\begin{figure}
    \centering
    \includegraphics[width=0.85\textwidth]{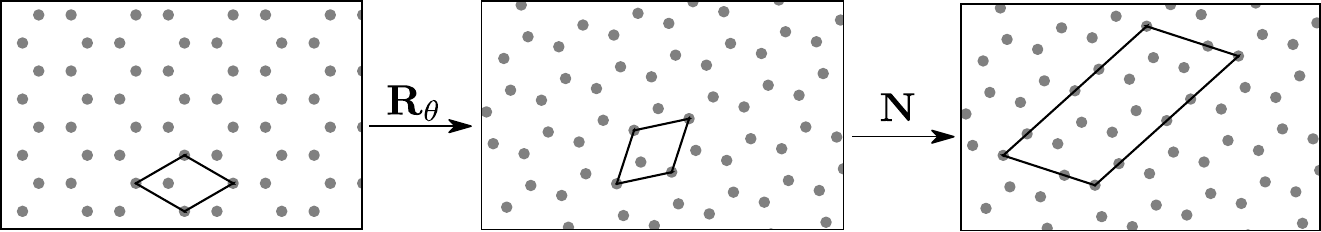}
    \caption{Illustration of going from primitive graphene cell to a rotated supercell. $\theta = 42^\circ$ radians and $(n_{00}, n_{01}, n_{10}, n_{11}) = (1,2,-2,2)$.}
    \label{fig:illust}
\end{figure}

% PB version
The mathematics of interfacing two graphene grains with a GB is closely related to how moiré cells are constructed \cite{necio2020supercell}, with the notable simplification that only one lattice vector needs to be common between the two sides of the GB. Having this condition fulfilled allows the construction of a GB as sketched in Fig. \ref{fig:GrainIllust}B and now formalise how to satisfy this condition in a simple way for two graphene lattices with a relative rotation and strain. Let the unit vectors of the hexagonal cell of the primitive graphene unit cell be $\mathbf{a}_1$ and $\mathbf{a}_2$, let $\mathbf{A} = [\mathbf{a}_1, \mathbf{a}_2]$, let $\mathbf{M},\mathbf{N} \in \{\mathbf{S}\in\mathcal{Z}^{2\times 2} \vert \mathrm{det}(\mathbf{S})\neq0\}$, let $\mathbf{R}_\theta$ be a rotation in 2D, $\epsilon$ a scalar strain, and search for solutions to 
\begin{align}
  \label{eq:system1}
  \mathbf{M} &= \left[ {\begin{array}{cc}
    m_{00} & m_{01} \\
    m_{10} & m_{11} \\
  \end{array} } \right]
  \phantom{MM}
  &&\mathbf{N} = \left[ {\begin{array}{cc}
    n_{00} & n_{01} \nonumber\\
    n_{10} & n_{11} \\
  \end{array} } \right]
  \phantom{MM} \nonumber \\ 
  \mathbf{A}_1 &= \mathbf{M}\mathbf{A}\phantom{MM}
  &&\mathbf{A}_2 = (1+\epsilon)\mathbf{N} \mathbf{R}_\theta\mathbf{A}\nonumber\\
  &\phantom{MMKKKKMMMM}\min_{\mu,\nu}\{|(\mathbf{A}_2)_\mu - (\mathbf{A}_1)_\nu|\}= 0\\
  &\phantom{MMMMMMMMM}\mu = 0 \phantom{m}\mathrm{or}\phantom{m}1, \quad \nu= 0 \phantom{m}\mathrm{or}\phantom{m}1 \nonumber
\end{align}
%\MB{Looks like python inside minimum and 1, 3rd the same?
%$\min_{\alpha,\beta}\{|(\mathbf{A}_2)_\alpha - (\mathbf{A}_1)_\beta|\}<t$}\ABL{I think this is normal math way of writing things, changed : to a dot instead. Corrected item 3 in the set.  }
The action of the rotation matrix $\mathbf{R}_\theta$ and the supercell matrices $\mathbf{M}$ and $\mathbf{N}$ is seen in Fig. \ref{fig:illust}.
Once two matching lattices with a small tolerable strain have been found, we construct an initial guess for the atomic coordinates. When the solution is found it is trivial to order the coordinate system so that the $y$-direction is along the GB, while the two lattice vectors describing the graphene periodicity on the left and right hand side of the GB are freely chosen from the rotated primitive lattice vectors. Methods for constructing GBs have also been considered previously and have then been relaxed using various potentials \cite{liu2011structure, zhang2013structures, ophus2015large}. 
However, here we furthermore perform a relaxation using DFT to achieve more realistic GB models. The steps of the workflow devised for this work are listed in Table 1:
\begin{enumerate}
\item[]
\fbox{
\parbox{\linewidth}{
\begin{enumerate}
    \item[] \phantom{mmmmmmmmmmmmmmmmm} \underline{Table 1}
    \item[0.] Find Solutions using random initialisation (2D)
    \begin{itemize}
        \item Loop over possible $\mathbf{M}, \mathbf{N}$, and by trial find a good initial guess for $\theta$.
        \item Solve eqs. \eqref{eq:system1} by a numerical solver to obtain $(\mathbf{M}, \mathbf{N},\theta,\epsilon)$. 
    \end{itemize}
    \item[1.] Geometry Creation (2D)
    \begin{itemize}
    \item Repeat one cell to the left, repeat the other cell to the right. Remove atoms that are within $0.85$ times the carbon-carbon bond length $d_{CC}$ to other atoms. This length has been chosen ad-hoc. 
    \item Place atoms at reasonable distances from each other, removing atoms that are too close and adding atoms where there are holes to be filled. A computationally cheap cost function $V$ (see SM.\ref{sec:Costfunction}) depending only on distances and angles to the surrounding neighbor atoms is used. An atom can be inserted if $V(\{\mathbf{r}_i \} \cup \{\mathbf{r}_{new}\})<V(\{\mathbf{r}_i \})$.
    Atoms further than $\Delta=8$Å away from the GB where the two flakes meet are considered as frozen and are not considered further in the calculation. 
    \end{itemize}
    \item[2.] Force field relaxation (3D)
    \begin{itemize}
    \item Geometry refinement with force-field relaxation using GULP\cite{gale2003general} and the Brenner potential\cite{Brenner_2002} for the atoms within $\Delta$ of the GB.
    \item If more than 2 counts of bond-angles are outside the interval of $[\theta_{min},\theta_{max}]=[100^\circ,160^\circ]$ the calculation is discontinued, and a new calculation is instead started from step 0. This interval is introduced ad hoc so that most bonds are closer to the $120^\circ$ bond angle of graphene.
    \end{itemize}
    \item[3.] Geometry relaxation using DFT (3D)
    \begin{itemize}
    \item Spin-degenerate DFT relaxation using the SIESTA code\cite{garcia2020siesta} of the atoms within $\Delta$ of the GB. Atoms outside this region are moved as a single, internally fixed structure on each side. Single-zeta basis is used and forces are required to be below $0.01$eV/\textup{~\AA}.
    \end{itemize}
    \item[4.] Final geometry relaxation using DFT-NEGF (3D)
    \begin{itemize}
    \item Spin-degenerate DFT relaxation using TranSIESTA\cite{papior2017improvements} which employs open boundary conditions and avoids crosstalk between neighboring GBs in a calculation with periodic boundary conditions. A single-zeta polarized basis is used and forces are required to be below $0.01$eV/\textup{~\AA}.
    \end{itemize}
\end{enumerate}
}
}
\end{enumerate}

\begin{figure}
    \centering
    \includegraphics[width=0.55\linewidth]{ 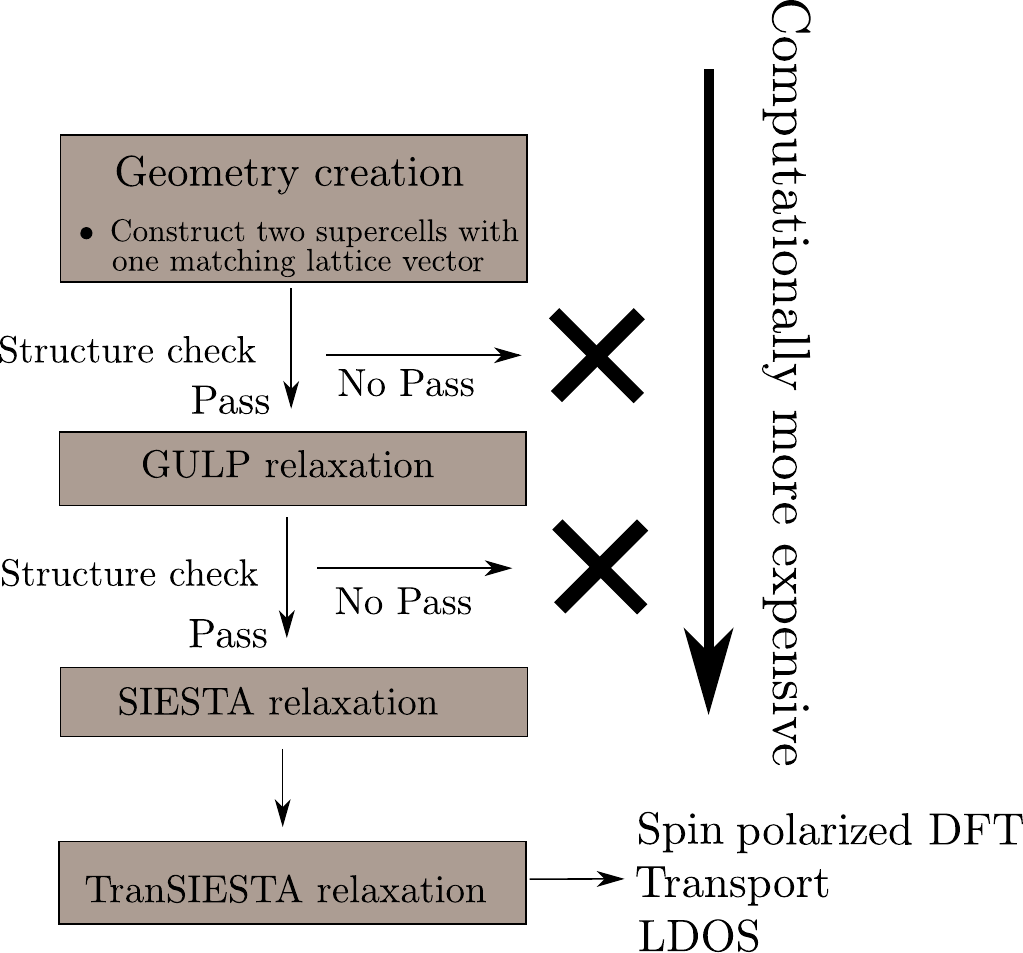}
    \caption{Flowchart of the workflow outlined in Table 1 for GB construction.}
    \label{fig:Flow}
\end{figure}

\begin{figure}
\centering
  \includegraphics[width=0.85\linewidth]{ 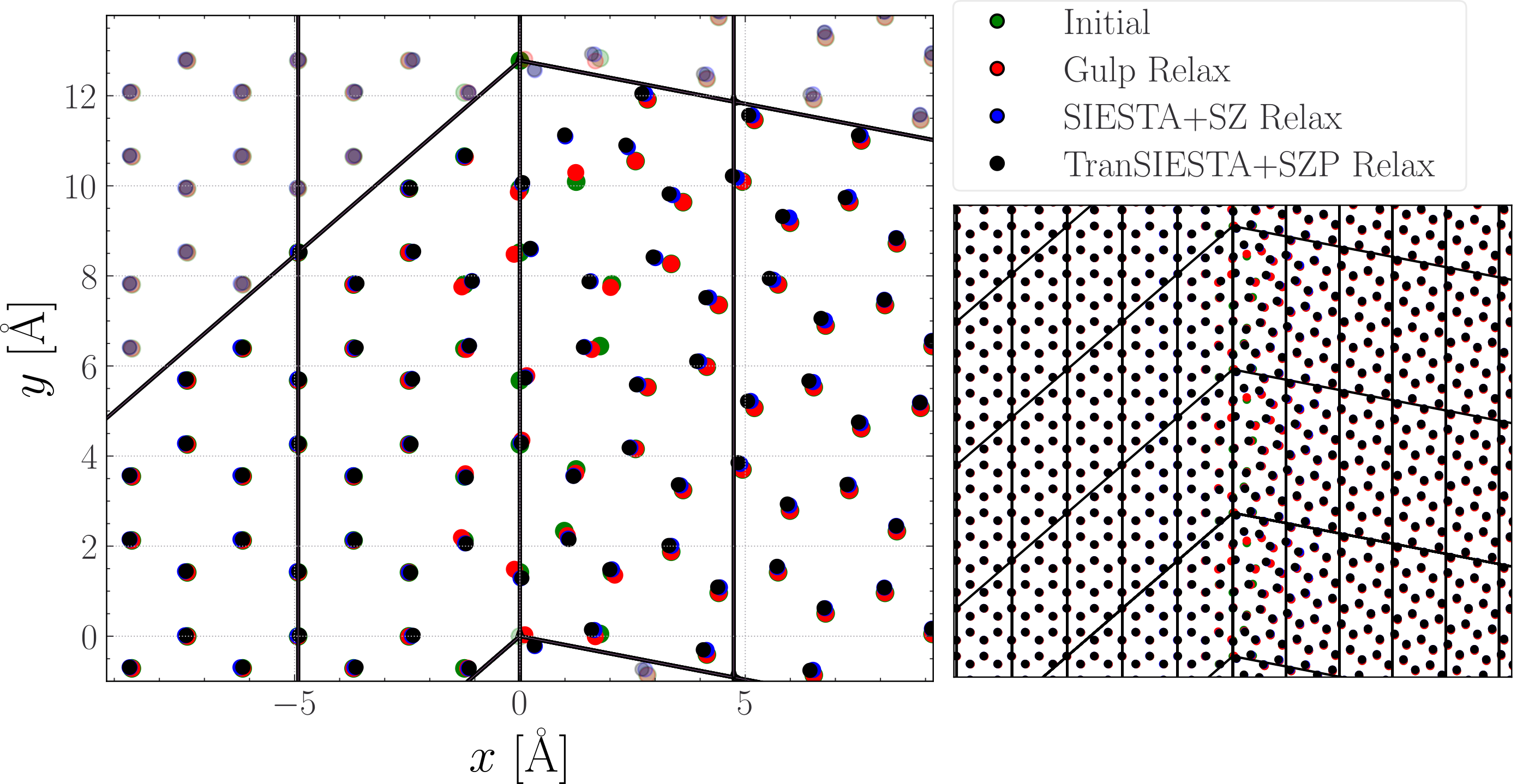}
  \caption{The individual steps of refinement of the workflow from Table 1 performed after solving the problem outlined in eq. \eqref{eq:system1}. Black lines indicate the unit-cell of the pristine graphene sheet on either side of the GB.}
  \label{fig:Fig1}
\end{figure}

This workflow is also visualized in Fig. \ref{fig:Flow} which shows the intermediate steps.
In Fig. \ref{fig:Fig1} an example of an initial structure is shown together with the intermediate and final steps. It can also be seen that the atoms close to the GB move substantially, while the atoms further away move in a rigid way. 
In the end of the workflow, the resulting GB may have relatively large out-of-plane buckling. A GB buckling in the $z$-direction can be a possibly sound configuration, but since the relaxing geometry has been constrained to be flat when $|x| > \Delta$, any in-plane and out-of-plane distortions are restricted to the region with $|x| < \Delta$.

\begin{figure}
\centering
  \includegraphics[width=0.65\linewidth]{ 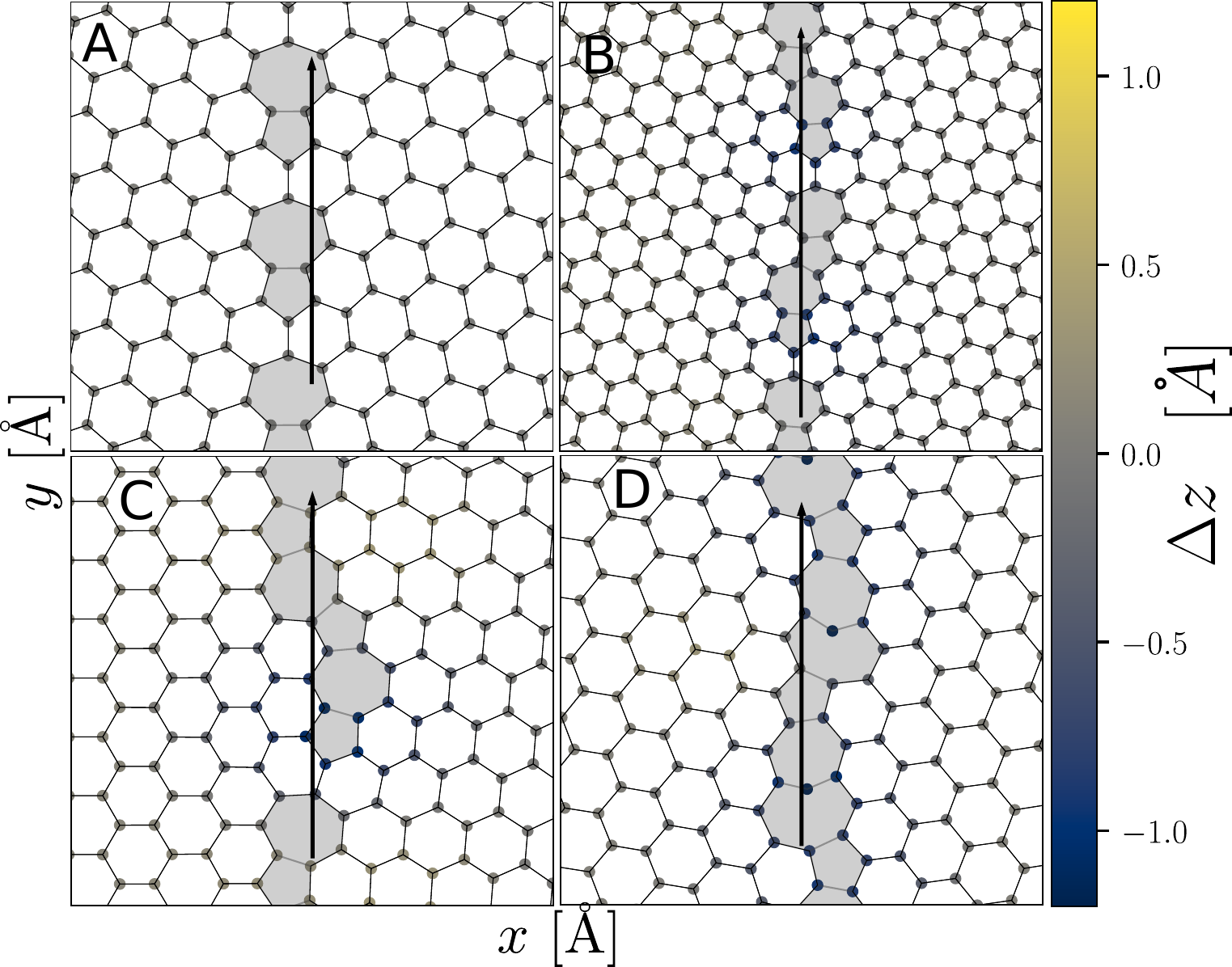}
  \caption{Examples of GBs with varying period and twist-angle. The structures are obtained using the workflow in Table 1. Pristine graphene  semi-infinite electrodes are attached to the left and right. A) The commonly found 5-7 GB with $\theta=21.79^\circ$. B) A variation of the 5-7 GB with $\theta=26.00^\circ$. C) Long GB with out-of-plane buckling and with $\theta=25.28^\circ$.D) Long GB containing two buckling carbon atoms and with $\theta=17.89^\circ$. Different scales in each subfigure. } 
  %A) The commonly found 5-7 GB. $\theta=21.79^\circ$. B) The commonly found zig-zag GB. $\theta=21.79^\circ$. C) Long GB with out-of-plane buckling $\theta=16.42^\circ$.D) Long GB containing a square. $\theta=9.44^\circ$. }
  \label{fig:4FinalStructures}
\end{figure}

%Experimental data suggests the widths of the deformation around a GB is roughly on the order of ~2 nm but varies.\cite{nemes2011revealing}\textit{[Remove this sentence?]} 
A selection of the relaxed structures coming out of the workflow can be seen in Fig. \ref{fig:4FinalStructures}, where the out-of-plane buckling also has been defined as $\Delta z=\max_i\{z_i - z_{avg}\}$. Furthermore, the initial structures have been chosen such that the length of the period of the GB is smaller than a maximal value $L_{GB}^{max}$ in order to restrict the required computational resources. 
%The hpc cluster at DTU has been utilized for the %workflow.\cite{DTU_DCC_resource}
% We can put \cite{DTU_DCC_resource} in the acknowledgements. 
%A 82 B 60 C 92 D 52 
% Case 1: 150
% Case 2: 190
% Case 3: 195
% Case 4: 77
\subsection{Non-equilibrium Greens Functions and Electronic Transport}
The theoretical framework for the DFT-based structural relaxation and electronic structure is the non-equilibrium Green's function framework implemented in the TranSIESTA code\cite{papior2017improvements}. The central object in this framework is the retarded Green's function, expressed in the device orbital basis, is given as\cite{papior2017improvements, haug2008quantum, stokbro2003transiesta}
\begin{align}
    \mathbf{G}^r_{k_y}(z) = (z\mathbf{S}_{k_y} -\mathbf{H}_{k_y} - \mathbf{\Sigma}^r_{k_y}(z))^{-1}.
    \label{eq:GF}
\end{align}
The Green's function is at $z = E + i\eta$ (for a numerically small $\eta$) from the device overlap matrix $\mathbf{S}_{k_y}$, device Hamiltonian $\mathbf{H}_{k_y}$, electrode self-energies $\mathbf{\Sigma}^r_{k_y}(z) = \sum_\alpha \mathbf{\Sigma}^r_{\alpha,k_y}(z)$ corresponding to the open boundary conditions in the non-periodic directions. 
A general matrix in $k_y$-space $\mathbf{C}_{k_y}$ can be written as a Bloch sum
\begin{align}
\label{eq:Blochsum}
     \mathbf{C}_{k_y} = \sum_{n}\mathbf{C}_{n}\mathrm{exp}[-2\pi i n k_yR_y],
\end{align}
where $\mathbf{C}_n$ is the collection of hopping matrix elements from the zero'th to the $n$'th unit cell. In the LCAO basis, this is the general prescription for how to calculate a matrix in $k_y$-space for a structure periodic in the $y$-direction. It is also useful, cf. Eq.~\eqref{eq:Blochsum} to define the unitless $\tilde{k}_y =  k_yR_y\in (-0.5,0.5]$. The energies at which $\mathbf{G}^r_{k_y}(z)$ has poles corresponds to states localized around the GB and can be found by solving the equation
\begin{align}
    \tilde{\mathbf{H}}_{k_y}(\epsilon_{k_y}^i)\psi_{k_y}^i = \epsilon_{k_y}^i \mathbf{S}_{k_y}\psi_{k_y}^i, \quad\mathrm{with}\quad \tilde{\mathbf{H}}_{k_y}(\epsilon)=\mathbf{H}_{k_y} + \mathbf{\Lambda}_{k_y}(\epsilon)\quad \mathrm{and}\quad 
    \mathbf{\Sigma}^r_{k_y} = \mathbf{\Lambda}_{k_y} - \frac{i}{2}\mathbf{\Gamma}_{k_y}.
    \label{eq:QP}
\end{align}
Equation \eqref{eq:QP} looks like the quasi-particle equation in many-body theory, but there are no many-body contributions to the self-energy in this case. The life-time instead comes from the finite size of the simulation region. We will call the states obtained by solving eq. \eqref{eq:QP} localized states (LS) from now on. Furthermore, the matrix element 
\begin{align}\label{eq:Lifetime}
    \bra{\psi_{k_y}^i}\mathbf{\Gamma}_{k_y}(\epsilon_{k_y}^i)\ket{\psi_{k_y}^i}\equiv \hbar/\tau_{k_y}^i
\end{align}
determines the life-time $\tau$. The state $\psi_{k_y}^i$ is an approximate eigenstate of the system Hamiltonian with a LS half life $\tau_{k_y}^i$. 
%At equilibrium, the QP states are also filled to the chemical potential.\cite{papior2017improvements}
\\
The self-energies $\mathbf{\Sigma}^r_{\alpha,k_y}$ are calculated from a bulk electrode Hamiltonian using a recursive algorithm for the calculation of the surface Green's functions of the electrodes\cite{sancho1985highly}. This algorithm is implemented in e.g TBtrans or sisl codes\cite{zerothi_sisl, papior2017improvements}.
%implemented e.g. in \hyperlink{https://github.com/zerothi/sisl}{sisl}.\cite{sancho1985highly, zerothi_sisl}
Given that the GBs considered here are periodic in the transversal $y$-direction, the real-space self-energy in a single cell $\mathbf{\Sigma}_\mathcal{R}$ can furthermore be found by a $k_y$-integral of the device Green's function $\mathbf{G}^r_{k_y}$ over the transverse Brillouin zone (TBZ)\cite{papior2019removing} as, 
\begin{align}
    \mathbf{G}^r_{\mathcal{R}}(z) &= \frac{1}{2\pi} \int_{\mathrm{TBZ}} \mathbf{G}^r_{k_y}(z)\mathrm{d}k_y \\ \mathbf{\Sigma}_\mathcal{R}^r(z) &= \mathbf{S}_{\vec{0}}z - \mathbf{H}_{\vec{0}} - (\mathbf{G}^r_\mathcal{R})^{-1}.
    \label{eq:RSSE}
\end{align}
Further computational details can be found in the SM. subsection \ref{sec:RSSEcalc}. 

In the general case when having calculated the electrode self-energies, the electrode broadening matrices 
$\mathbf{\Gamma}_{\alpha, k_y}(z) = i\left[\mathbf{\Sigma}^r_{\alpha,k_y}(z)- \mathbf{\Sigma}^a_{\alpha,k_y}(z)\right]$ can be evaluated. This makes it possible to calculate the electronic transmission function from left to right as\cite{meir1992landauer}
\begin{align}
    T(E,k_y) &= \mathrm{Tr}[\mathbf{G}^{r}_{k_y}(E)\mathbf{\Gamma}_{L,k_y}(E)\mathbf{G}^a_{k_y}(E)\mathbf{\Gamma}_{R,k_y}(E)] \\
    T(E) &=\int_{-0.5}^{0.5} T(E, {\tilde{k}_y}) \,\mathrm{d}{\tilde{k}_y}.\label{eq:LandaurFormula}
\end{align}
Equation \eqref{eq:LandaurFormula} gives the elastic transmission probability accounting for the electron wave interfering with itself as it moves through the structure. The transmission function calculated with TBtrans will in following plots be normalized with respect to the associated GB period, $L_{GB}$.

Lastly, the spectral function $\mathbf{A}_{\alpha,k_y}$ can be calculated from the Green's function and the electrode broadening function, given as\cite{papior2017improvements}
\begin{align}
    \mathbf{A}_{\alpha,k_y}(z) = \mathbf{G}^r_{k_y}(z)\mathbf{\Gamma}_{\alpha, k_y}(z)\mathbf{G}^a_{k_y}(z).
\end{align}
It is useful for quantifying bound states in the system, as $\mathbf{A}_{\alpha,k_y}$ only contains the states which couple into electrode $\alpha$, and thus does not contain the contribution from the LS at the GB.
%The energies of interest taken to be between $E=-2.0$ eV and $E=2.0$eV since the Fermi level in graphene moved roughly $\pm 0.5$eV by a backgate and then an additional system bias might give another $\pm 1.0$eV in the case of a very significant system bias. 
%When calculating transmissions with either eq. \eqref{eq:LandaurFormula} or \eqref{eq:naivetransmission}, it is really the transmission per unit length of the GB that is of interest as GBs of different lengths are then directly comparable.
%\subsection{Diffusive Conductivity along the GB}

%\subsection{Inelastic Transport}
%To be written

\subsection{Density Functional Theory}
The final step of relaxing the structures and the transport calculations is performed using the PBE exchange-correlation functional\cite{perdew1996generalized}, an energy cutoff of 150Ry and a SZP basis set in TranSIESTA\cite{papior2017improvements}. A simulation of a system being doped e.g. by a back-gate or dopant atoms, is done using mixed pseudo-potentials, also known as the virtual crystal approximation\cite{bellaiche2000virtual}. Here a weighted sum of carbon and nitrogen or boron is used to add or remove charge from the system.
%%%%%%%%%%%%%%%%%%%%%%%%%%%%%%%%%%%%%%%%%%
\section{Results}
\subsection{Atomic Geometries}
\begin{figure}
\centering
\includegraphics[width=0.9\linewidth]{ 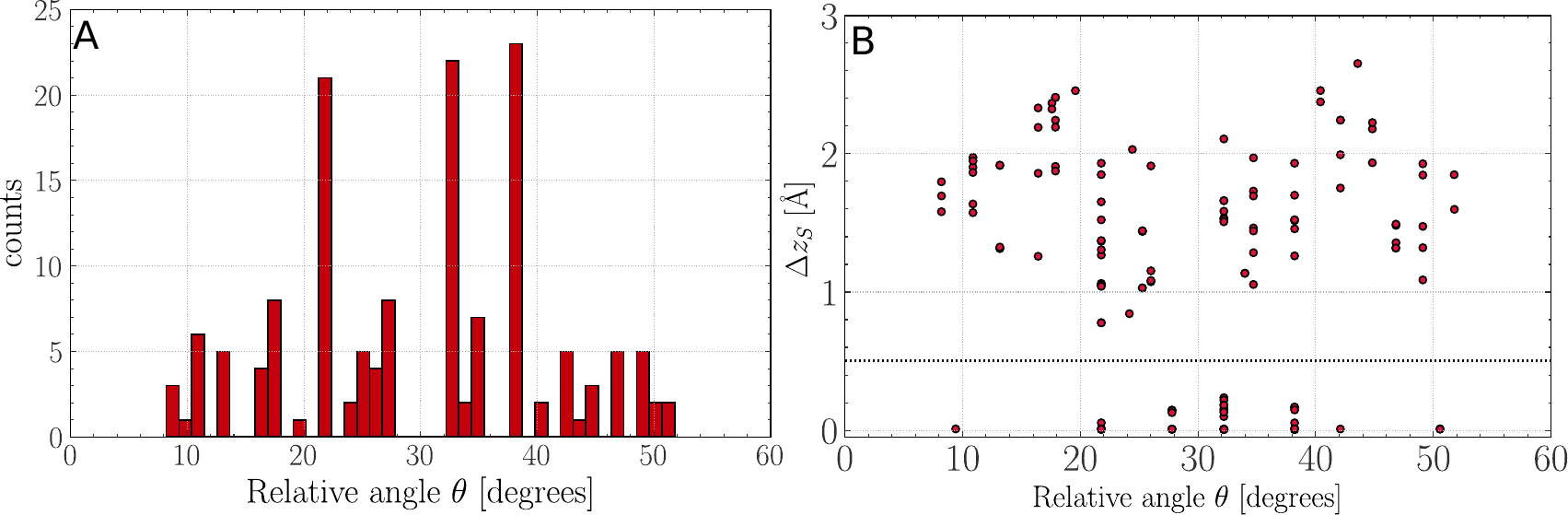}
%  \begin{subfigure}{0.35\linewidth}
%    \includegraphics[width=1\linewidth]{Pictures/Angle_histogram.png}
%    \caption{Histogram of the relative twist angle of the created geometries} \label{fig:AngularHist}
%  \end{subfigure}%
  %\hspace*{\fill}   % maximize separation between the subfigures
%  \begin{subfigure}{0.35\textwidth}
%    \includegraphics[width=1\linewidth]{Pictures/Angle_vs_deltaZ.png}
%    \caption{Scatter-plot of the out-of-plane buckling of the created geometries} \label{fig:bucklingscatter}
%  \end{subfigure}
%  \\[1ex]
%  \begin{subfigure}{0.35\textwidth}
%    \includegraphics[width=1\linewidth]{Pictures/Angle_vs_variance.png}
%    \caption{Scatter-plot of the variance of the $z$-component of the atomic placements of the created geometries within $|x|<10$Ang of the GB.} \label{fig:variancescatter}
%  \end{subfigure}%
\caption{Statistics of the dataset. A) Histogram of GBs at each angle. B) Maximal difference between highest and lowest (in $z$-direction) atoms in the GB structure.% C) Variance of the $z$-coordinate of atoms within $|x_i|<1$ nm of the GB.
}
\label{fig:1}
\end{figure}

The relative angle of rotation $\theta$ of the two lattices forming the periodic GB, together with the out-of-plane buckling are basic parameters that influence the conductive properties. A histogram of the angles contained in the data set is shown in Fig. \ref{fig:1}A. This data set spans a wide number of angles, but exhibits gaps, for example in the vicinity of $\theta=30^\circ$ where no GB structures were identified. 
%A scatter-plot of the maximal out-of-plane buckling and variance of the atomic $z$-displacements can be seen in \ref{fig:1}B and  Fig. \ref{fig:1}C. 
A scatter-plot of the maximal out-of-plane buckling of the atomic $z$-displacements can be seen in Fig. \ref{fig:1}B. 
From these figures it appears that a majority of the \TotalStructs GBs produced shows a significant buckling within the structure, while only \FlatStructs GBs are found  below the line at $\Delta z_S = 0.5$Å in Fig. \ref{fig:1}B. The significance of out-of-plane buckling is to enable hybridization between the $\pi$- and $\sigma$-subsystems that are decoupled in bulk graphene. This might also be the reason for the lack of structures around the line $\Delta z_S = 0.5$Å in Fig. \ref{fig:1}B, where the carbon atoms either tend to hybridize in a sp$^3$ configuration with buckling or by sp$^2$ in-plane bonding. 
Notably, the GB structures  below this line cluster around relatively few angles. 
%A majority of these is however also in the lower end of the variance plot in Fig. \ref{fig:1}C with \LowVarStructs GBs having a variance below 0.1Ang$^2$, meaning fewer atoms are sticking out.
The angles available in this data set are limited by the maximum GB length $L_{GB}^{max}$ since it is certainly possible to generate GBs at smaller angles, they just have a very long periodicity, thus requiring more computational resources.  
 %The method for generating these GBs are furthermore limited by the length of the periodic GBs, since it is certainly possible to generate GBs at smaller angles, they just have a very long periodicity, which is not computationally feasible to consider for our purposes. 
The whole data set presented in Fig. \ref{fig:1} is available in Ref. \cite{githubGB}.%at \MB{hyperlinks are normally not allowed/used inside scientific articles: References are used and then the hyperlink as text can be there.} \hyperlink{google.com}{LinkToSomewhere}. 

\subsection{Conductive Properties and Electronic Structure}
\begin{figure}[h]
\centering
\includegraphics[width=1.0\textwidth]{ 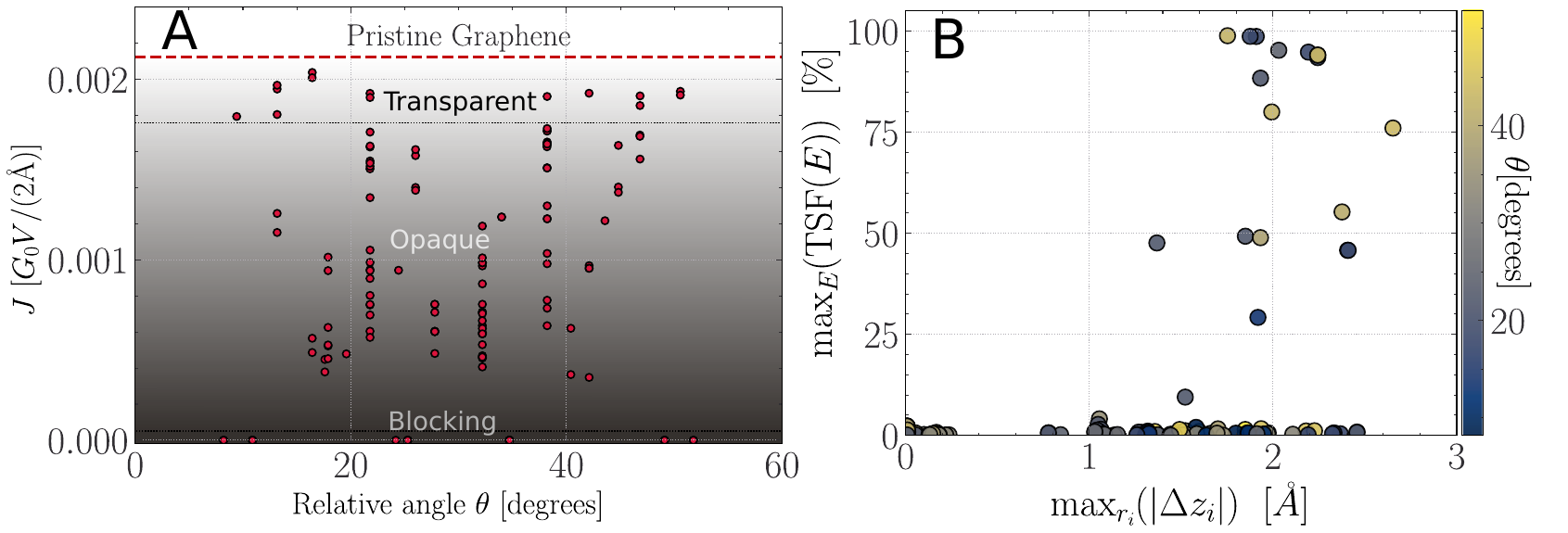}
%  \begin{subfigure}{0.35\textwidth}
%    \includegraphics[width=1\linewidth]{Pictures/3D_transmission.png}
%    \caption{} \label{fig:3D1}
%  \end{subfigure}
%  %\hspace*{\fill}   % maximize separation between the subfigures
%  \begin{subfigure}{0.35\textwidth}
%    \includegraphics[width=1\linewidth]{Pictures/3D_transmission_2.png}
%    \caption{} \label{fig:3D2}
%  \end{subfigure}
%  \\[1ex]
%  \begin{subfigure}{0.35\textwidth}
%    \includegraphics[width=1\linewidth]{Pictures/3D_transmission_3.png}
%    \caption{} 
%    \label{fig:3D3}
%  \end{subfigure}
% % \hspace*{\fill}
%  \begin{subfigure}{0.35\textwidth}
%    \includegraphics[width=\linewidth]{Pictures/3D_transmission4.png}
%    \caption{} 
%    \label{fig:3D4}
%  \end{subfigure}
%\transmission $K^{L\rightarrow R}(E)$,  while the shaded blue lines are the %$T^{L\rightarrow R}(E)$ functions for the GB. a) \& b) All GBs, two %different angles. c) \& d) GBs with $\Delta z_S< 0.5$Ang, two different %angles.  } 
%\label{fig:3D_transmission}
\caption{A) Current density across a GB at a bias $V = 0.2$V without including non-equilibrium effects. The current density $J = J_\uparrow + J_\downarrow$ is calculated using the equilibrium transmission function and a symmetric voltage drop, $V_R = -V_L = 0.1$V. B) Scatter-plot of the TSF against maximal buckling of the structure. Maximum taken over $E\in [-1/2, 1/2]$eV}
\label{fig:J_scatter}
\end{figure}

%How well these GBs allow for current conduction through them is of course of interest if one wishes to make devices, where the GBs, willingly or unwillingly, are present. 
The transmission of current across GBs is important for the sheet resistance of large-area polycrystalline graphene films and the performance (and variability of performance) of any devices made thereof. 
We consider here the transmission function $T(E)$ per unit length of GB. The Landauer-Büttiker formula determines the current density running across the GB at zero temperatures by\cite{papior2017improvements}
\begin{align}
    J_s= \frac{G_0}{2eL_{GB}}\int_{V_L}^{V_R}T_s(E)\mathrm{d}E, 
\end{align}
%\MB{why a factor-2 in the denominator? It is spin-degenerate and we have 2 there in $G_0$ to account for that.}\ABL{Its the same as in the "improvements" article?}
where $s$ is spin and $T(E)$ beyond linear response also depends on the applied voltage. $G_0=\frac{2e^2}{h}$ and $h$ is Planck's constant. The quantitative behavior of the current density relative to the twist angle is shown in Fig. \ref{fig:J_scatter}A and it displays a large variance of how well the GBs conduct current. Some GBs conduct current as well as graphene ("transparent regime"), while in other GBs, the current is numerically negligible ("blocking regime").  Yet more GBs are somewhere in between these limits ("opaque regime"). The categorisation in Fig. \ref{fig:J_scatter}A is of course subjective and the transition between opaque and transparent is smooth.
%, which is seen in Fig. \ref{fig:3D_transmission}.
The cause of the blocking behavior for some GBs is momentum mismatch between the available electrode states on each side as identified in ref. \cite{yazyev2010electronic}. The intermediate, "opaque regime", is interesting because a combination of available scattering states on either side and wave interference effects cause the transmission function to be significantly suppressed. 

%The naive transmission function is in its broad features similar to the pristine graphene  transmission function, with its characteristic "V"-shape. There is furthermore also a presence of regions where both $T$ and $K$ is zero, i.e. there are gaps where elastic transport is not allowed in some GBs. This has been covered before\cite{yazyev2010electronic} and is simply because these regions does not have a state on the other side of the GB, which also has the same transverse crystal momentum. This is also encoded in the naive transmission function $K$.\\
%The top row in Fig. \ref{fig:3D_transmission} furthermore shows that there are both GBs which transmit very well over a wide range of energies and GBs which suppress the flow of electrons through them significantly, even though there are states available to tunnel from and to on each side of the GB. This suppression of current is then coming from destructive interference of the electrons as it gets diffracted by the GB. The angles which this suppression current current happens at is more clear in the case where the dataset is restricted to the GBs with minimal buckling. Here the two groupings around $\theta=32^\circ$ and $\theta=28^\circ$ contributes the majority of cases where the transmission is strongly suppressed. The case is not so clear in the full dataset, where scattering off the buckling defects seem to also play a larger role. 
The current density in Fig.~\ref{fig:J_scatter}A is summed over spin, meaning spin-dependent effects are not visible. In Fig.~\ref{fig:J_scatter}B the spin filtering of the GBs in the data set is quantified though the Transmission Spin Filtering(TSF) coefficient, defined as
\begin{align}
    \mathrm{TSF} = \frac{|T_{\uparrow} - T_{\downarrow}|}{T_{\uparrow} + T_{\downarrow}}\times 100\%.
\end{align}

%The cases with the larger buckling, spin-polarised states also starts to emerge. Here we report the spin-filtering efficiency in terms of the in filtering fraction $\mathrm{SF}_{\epsilon}(E) = |\frac{T_{\uparrow}(E)-T_{\downarrow}(E)}{T_{\uparrow}(E)+T_{\downarrow}(E) + \epsilon}|$, where $\epsilon$ is a small positive number introduced to avoid large spin-filtering fraction when the transmission is numerically negligible. 
A plot of the maximal value of TSF$(E)$ with $-0.5\mathrm{eV}<E<0.5$eV for the various structures is seen Fig. \ref{fig:J_scatter}B, showing that out-of-plane buckling makes some GBs spin-polarize and can work as a spin-filter. %The largest spin-filtering effect is seen in the cases with the most buckling, but there are GBs with modest buckling that also spin-polarizes.
%\begin{figure}[h]
%\centering
 % \begin{subfigure}{0.35\textwidth}
 %   \includegraphics[width=\linewidth]{Pictures/SF1}
 %   \caption{$\Delta z_S<0.5$Ang. 48 counts.} \label{fig:SF1}
 % \end{subfigure}
  %\hspace*{\fill}   % maximize separation between the subfigures
 % \begin{subfigure}{0.35\textwidth}
 %   \includegraphics[width=\linewidth]{Pictures/SF2.png}
 %   \caption{$0.5<\Delta z_S<1.5$Ang. 38 counts.} \label{fig:SF2}
 % \end{subfigure}
 % \\[1ex]
 % \begin{subfigure}{0.35\textwidth}
 %   \includegraphics[width=\linewidth]{Pictures/SF3.png}
 %   \caption{$1.5<\Delta z_S<2.5$Ang. 49 counts.} 
 %   \label{fig:SF3}
 % \end{subfigure}
  %\hspace*{\fill}
 % \begin{subfigure}{0.35\textwidth}
 %   \includegraphics[width=\linewidth]{Pictures/SF4.png}
 %   \caption{$\Delta z_S>2.5$Ang. 1 count.} 
 %   \label{fig:SF4}
 % \end{subfigure}
%\caption{Spin-filtering fraction SF$_{0.01}(E)$ plotted for various intervals of $\Delta z_S$. } 
%\label{fig:SFplot}
%\end{figure}
In the most extreme case the TSF comes out at close to $\sim 100\%$ in Fig. \ref{fig:J_scatter}B, indicating a GB that completely polarizes the current. %\MB{??3:1 well you get then $(3-1)/(3+1) = 0.5$, no?}
%Furthermore, some of the GBs the preferred spin to transmit through the GB also changes depending on where the Fermi-level is located. This makes these GBs a gate-tuneable gate spin-filter that can tuned to transmit either spin preferably, within this theory.
We can identify four overall types of GBs in this dataset, which are summarized in Table 2. 
\begin{enumerate}
\item[]\hspace{-.75cm}
\fbox{
  \parbox{0.98\linewidth}{
\begin{enumerate}[label=\textnormal{(\arabic*)}=]
    \item[]\phantom{mmmmmmmmmmmmmmmmm} \underline{Table 2}
    \item[1.] Well-transmitting GBs $J>0.9J_{pristine}$ c.f Fig. \ref{fig:J_scatter}, which do not scatter the electrons significantly. These are mostly flat GBs with twist-angle around $\theta \approx 38^\circ$ and $\theta \approx 21^\circ$. 
    \item[2.] Opaque GBs $0.9J_{pristine}>J>0.05J_{pristine}$ c.f Fig. \ref{fig:J_scatter} that transmit poorly due to electron diffraction through the GB, and are mostly present around angles $\theta \approx 28^\circ$ and $\theta \approx 21^\circ$. 
    %A subset of these GBs exhibit significant destructive interference at particular energies, giving lowered conduction even as the electron concentration in the sample is changed. 
    \item[3.] Blocking GBs $J<0.05J_{pristine}$ with transport gaps arising from mismatch in crystal momentum\cite{yazyev2010electronic}.
    \item[4.] Spin-filtering GBs that spin-polarize and transmit a preferred spin. This case can fall into any of the previous three categories.
\end{enumerate}
}
}
\end{enumerate}

\begin{figure}
  \includegraphics[width=1.0\linewidth]{ 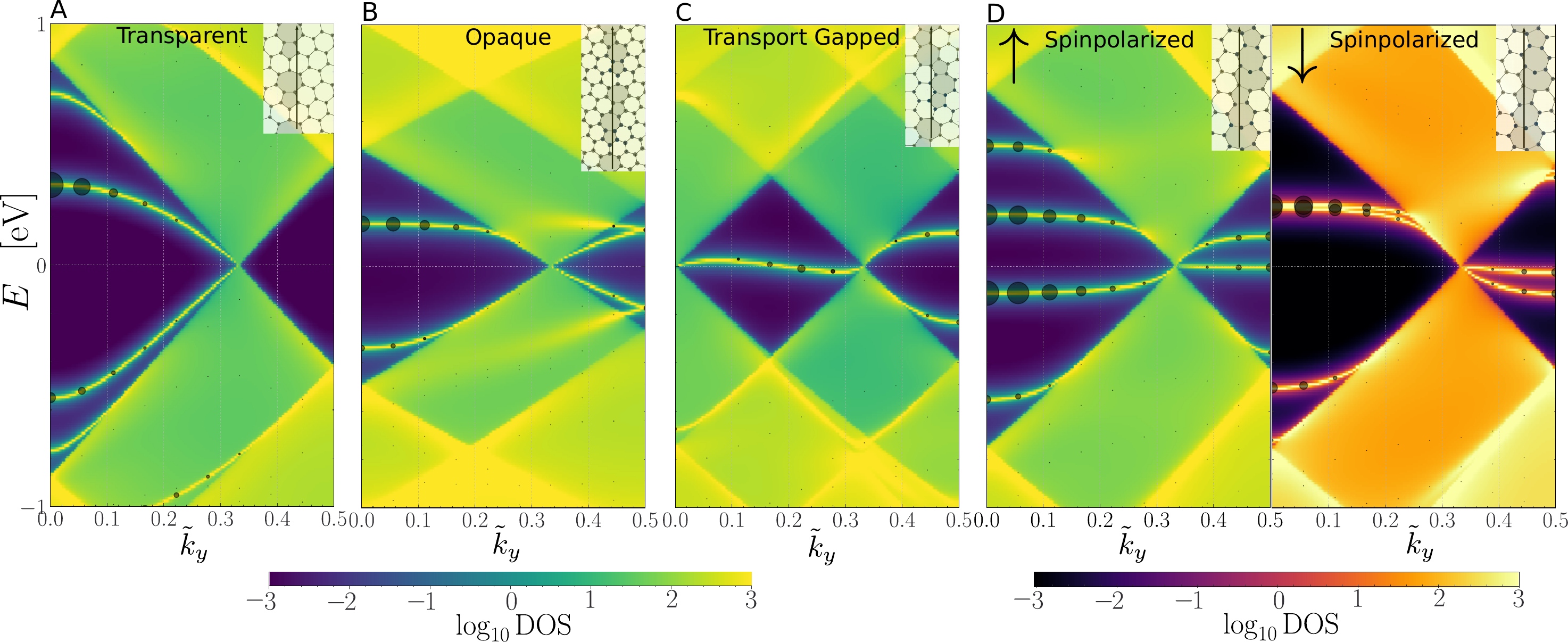}
  \caption{ A-D) GB DOS as a function of energy $(E)$ and transverse wavevector $(k_y)$ for the GBs shown in Fig. 3A-D, respectively. The size of the black circles scale with the lifetime $\tau_{k_y}^i$ of the LS (largest bubble  corresponds to $\approx0.5$ns). In D) the two panels are for the separate spins of the spin-polarized GB. Note the different color scheme used for spin down in the fifth panel from the left. The color bar associated with the second color-scheme is the second from the left. }
  \label{fig:2D_DOS}
\end{figure}

In Fig. \ref{fig:2D_DOS}A/B/C/(D+E) the electronic structure of the GBs in Figs. \ref{fig:4FinalStructures}A-D is characterised by the  DOS obtained from the imaginary part of the Green's function. The DOS is resolved in both energy and transverse $k_y$-point in the TBZ, and the plot also shows the LS lifetime $\tau^i_{k_y}$ from eq. \eqref{eq:QP}. 
The Dirac cone of graphene is clear in all five panels of Fig. \ref{fig:2D_DOS}. There are, however, also other band-like features. These come from LS around the GB. These states act as a separate subsystem with its own electronic structure and they couple very weakly to the bulk graphene electrode states. %We will call these defect states for quasi-particle states.
The structures shown in Figs. \ref{fig:4FinalStructures}A-D fall into cases 1-4 of Table 2, respectively. 
Ayuela {\it et al.}\cite{ayuela2014electronic} used tight-binding calculations to show how the number of LS bands depends on the edge-states of the two semi-infinite graphene sheets that connect via the GB, and thus classified these in terms of the edges of either side. In this work we instead choose to describe the GBs in terms of their conductive properties cf. Fig. \ref{fig:2D_DOS} and Table 2.

We may characterize the localized GB states using the quasi-particle formalism and associate a life-time cf. eqs.~\eqref{eq:QP} and \eqref{eq:Lifetime}. 
The intrinsic LS lifetime in graphene has been found in experiments to be above  $\sim0.5$ ps corresponding to $\eta \approx 10$meV and a mean free path on the order of 100nm\cite{li2009scanning}. In our calculation the LS lifetime is an artefact due to the truncation of the LS in the finite device region between the electrodes (Always $>$40Å until the electrodes start on either side of the GB). The very long lifetime we obtain in the calculation (above $\sim 0.5$ns) compared to intrinsic graphene means that we should consider these as localized. On the other hand we may use the relative change in the computed lifetimes as a measure of their decay and overlap with the electrode regions. With this in mind, we see how the lifetime (circles in Fig. 6) grow with distance to the electrode states in the Dirac cone.
%\MB{I would say that this life-time is "artificial" since it depends on out numerical choices of the size of electrode region -- distance from GB to electrode (and maybe $\eta$). In an exact/ideal calculation the DOS for QP states would be delta-functions?? }
%\MB{Suggestion for text: The intrinsic quasi-particle life time in graphene has been found in experiments to be above $\sim 0.5$ps corresponding to $\eta\approx 10$meV and a mean free path on the order of 100nm DOI10.1103/PhysRevLett.102.176804. In our calculation the QP life time is an artefact due to the truncation of the localized defect states in the finite device region between the electrodes. The very long life time we obtain in the calculation (above $\sim 0.5$ns) compared to intrinsic graphene means that we should consider these as localized. On the other hand we may use the relative change in the computed life times as a measure of their decay and overlap with the electrode regions. With this in mind, we see how the life time (circles in  Fig. \ref{fig:2D_DOS}) grow with distance to the electrode states in the Dirac cone.}  
In Fig. \ref{fig:2D_DOS}A the bands merge with the Dirac cone as they get close to the projection of the graphene $K$-point on the GB direction. This is, however, not the case in Fig.~\ref{fig:2D_DOS}B where  signatures of the bands associated with the LS can also be observed inside the region of the Dirac cone. A similar situation appears in Fig.~\ref{fig:2D_DOS}D and Fig.~\ref{fig:2D_DOS}E. 

\begin{figure}
  \includegraphics[width=1.0\linewidth]{ 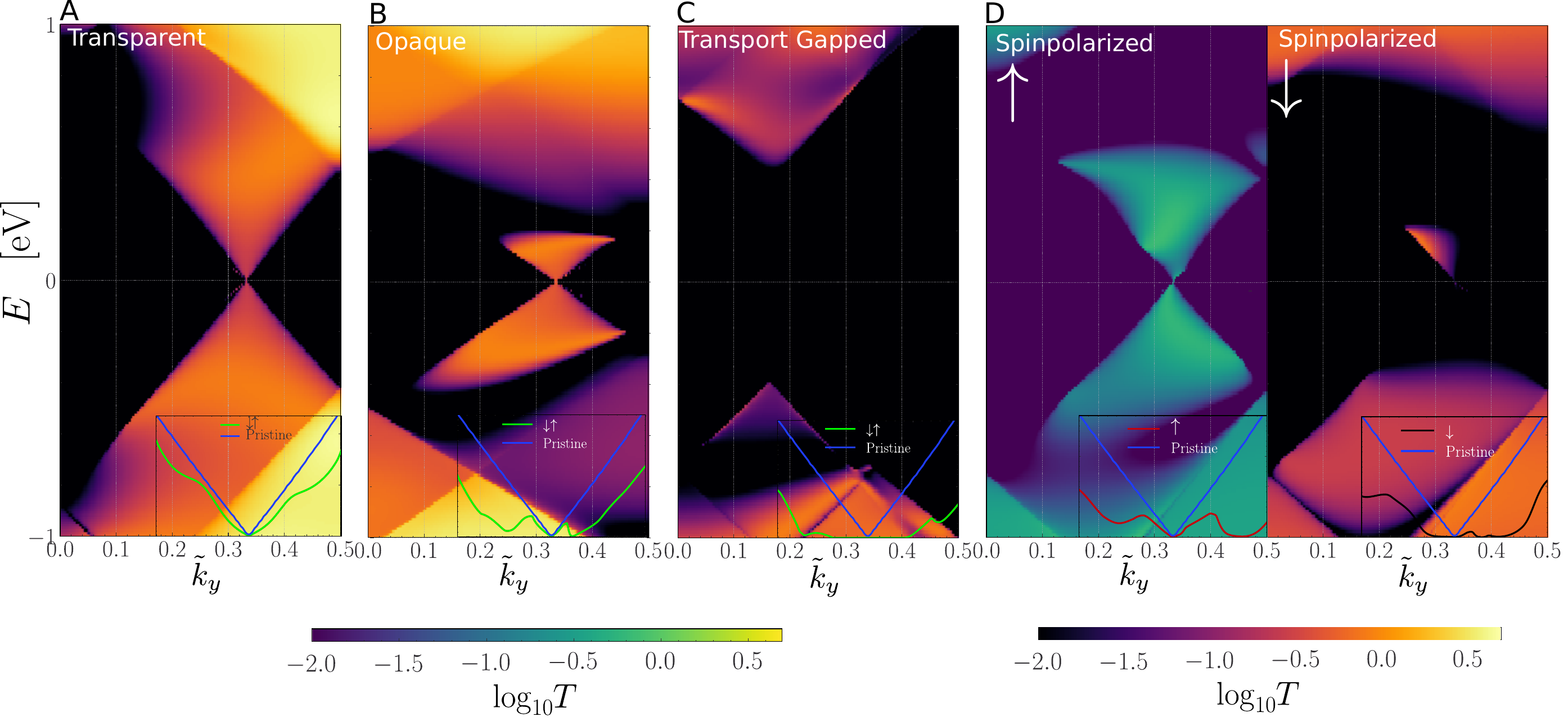}
  %\caption{$E$- and $k_y$-resolved GB transmission function. A) $T(E,k_y)$ of Fig. \ref{fig:4FinalStructures}A B) $T(E,k_y)$ of Fig. \ref{fig:4FinalStructures}B C) $T(E,k_y)$ of Fig. \ref{fig:4FinalStructures}C D) $T(E,k_y)$ for spin-up of Fig. \ref{fig:4FinalStructures}D E) $T(E,k_y)$ for spin-down of Fig. \ref{fig:4FinalStructures}D.  A broadening $\eta=10^{-3}$eV has been used. The inset in each figure is the $k_y$-integrated transmission function where the first axis is energy and second axis is $T(E) / L_{y}$. A broadening $\eta=10^{-3}$eV has been used.}
  \caption{A-D) Transmissions  as a function of energy ($E$) and transverse wavevector ($k_y$) for the GBs shown in Fig. 3 A-D, respectively, representing the transparent, opaque, gapped, and spin-polarized cases. A broadening $\eta=10^{-3}$eV has been used.}
  \label{fig:2D_Tr}
\end{figure}

\begin{figure}
  \includegraphics[width=1\linewidth]{ 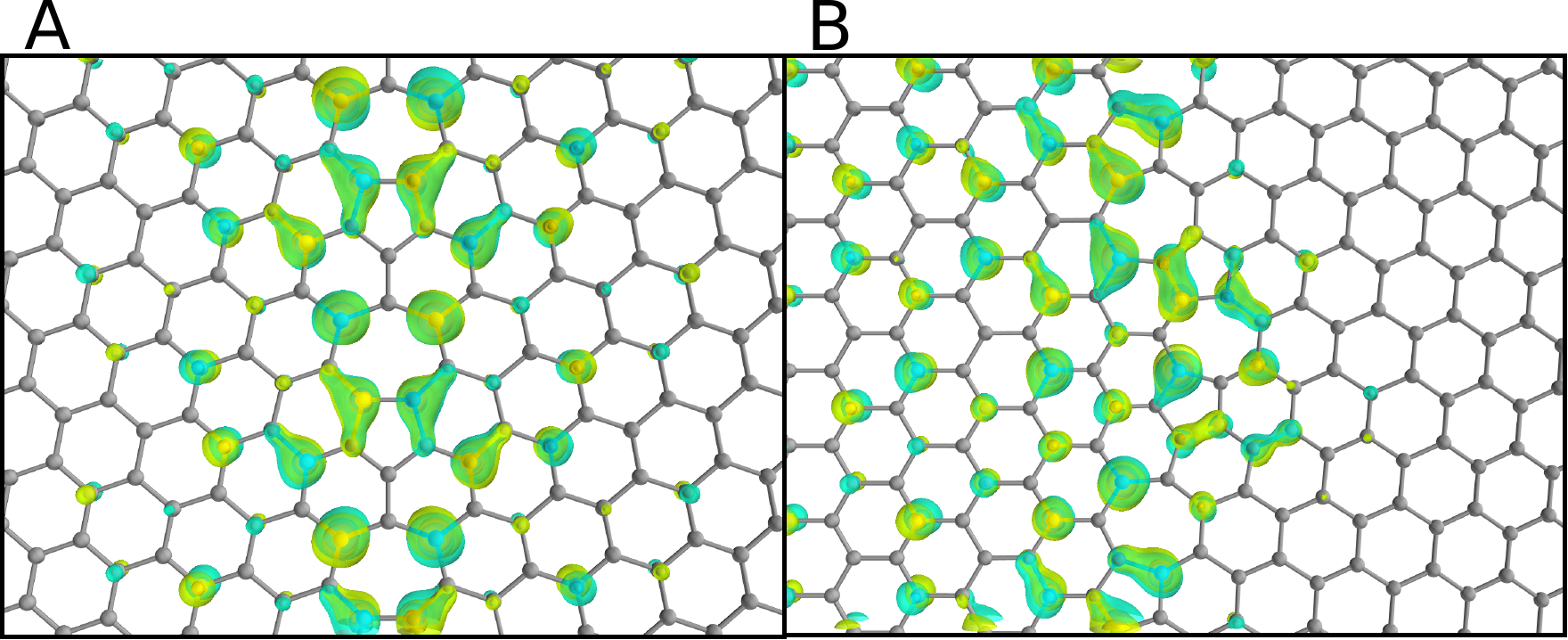}
  \caption{A) LS of first conduction band of Fig. \ref{fig:2D_DOS}A at $k_y = 0$ and $E=0.33$eV. The lifetime of this state is $\tau \approx 0.28$ns. B) LS of the flat band in Fig. \ref{fig:2D_DOS}C at $k_y = \frac{\pi}{3L_{GB}}$ and $E = 0.01$eV. The lifetime of this state is $\tau \approx 0.011$ns. Length-scales in A) and B) are different.}
  \label{fig:QPstate}
\end{figure}

The effects of the LS bands intersecting the Dirac cone are clear from $T(E, k_y)$, shown in Fig. \ref{fig:2D_Tr}.
The presence of the LS bands facilitates transmission at the intersection point, but also gives rise to a sharp decline when moving above/below the LS bands, see Fig. \ref{fig:2D_Tr}B-D. Lastly, the LS of the first conduction band of the GB in Fig. \ref{fig:2D_DOS}A is plotted in Fig. \ref{fig:QPstate}A.
The GB is symmetric around $x=0$ and the state shown in this figure has negative parity i.e. it changes sign across the GB. In Fig. \ref{fig:QPstate}B the LS of the flat band in Fig. \ref{fig:2D_DOS}C is shown. This state is instead skewed towards one of the electrodes with a longer decay length and thus stronger truncation, which explains the shorter lifetime of the flat-band in Fig. \ref{fig:QPstate}B compared to the state in Fig. \ref{fig:QPstate}A. Fig. \ref{fig:2D_DOS}C furthermore shows a peculiar situation with a different number of bands on either side of the $k_y = 1/3$ point. In all DOS-plots, except Fig. \ref{fig:2D_DOS}C, the Dirac cones of either side of the GB are on top of each other, while in Fig. \ref{fig:2D_DOS}C the geometry is such that the Dirac cone is centered at different $k_y$-points, which shows how this case is momentum-mismatched yielding a transport gap in Fig.~\ref{fig:2D_Tr}C.
The bands seen in Fig.~\ref{fig:2D_DOS} give rise to a DOS that is significantly modified relative to pristine graphene, with the notable introduction of van Hove singularities in the DOS where the LS bands are flat. 
This behavior has for instance been observed in ref. \cite{luican2016localized,ma2014evidence} by voltage-dependent differential conductance ($\frac{\mathrm{d}I}{\mathrm{d}V}$) measurements by STM-spectroscopy on top of the GB.
%This has for instance been observed in ref. \cite{luican2016localized,ma2014evidence} by $\frac{\mathrm{d}I}{\mathrm{d}V}$ measurements by STM-spectroscopy on top of the GB. 

\subsubsection{Non-equilibrium effects}

\begin{figure}
    \centering
    \includegraphics[width=.75\linewidth]{ 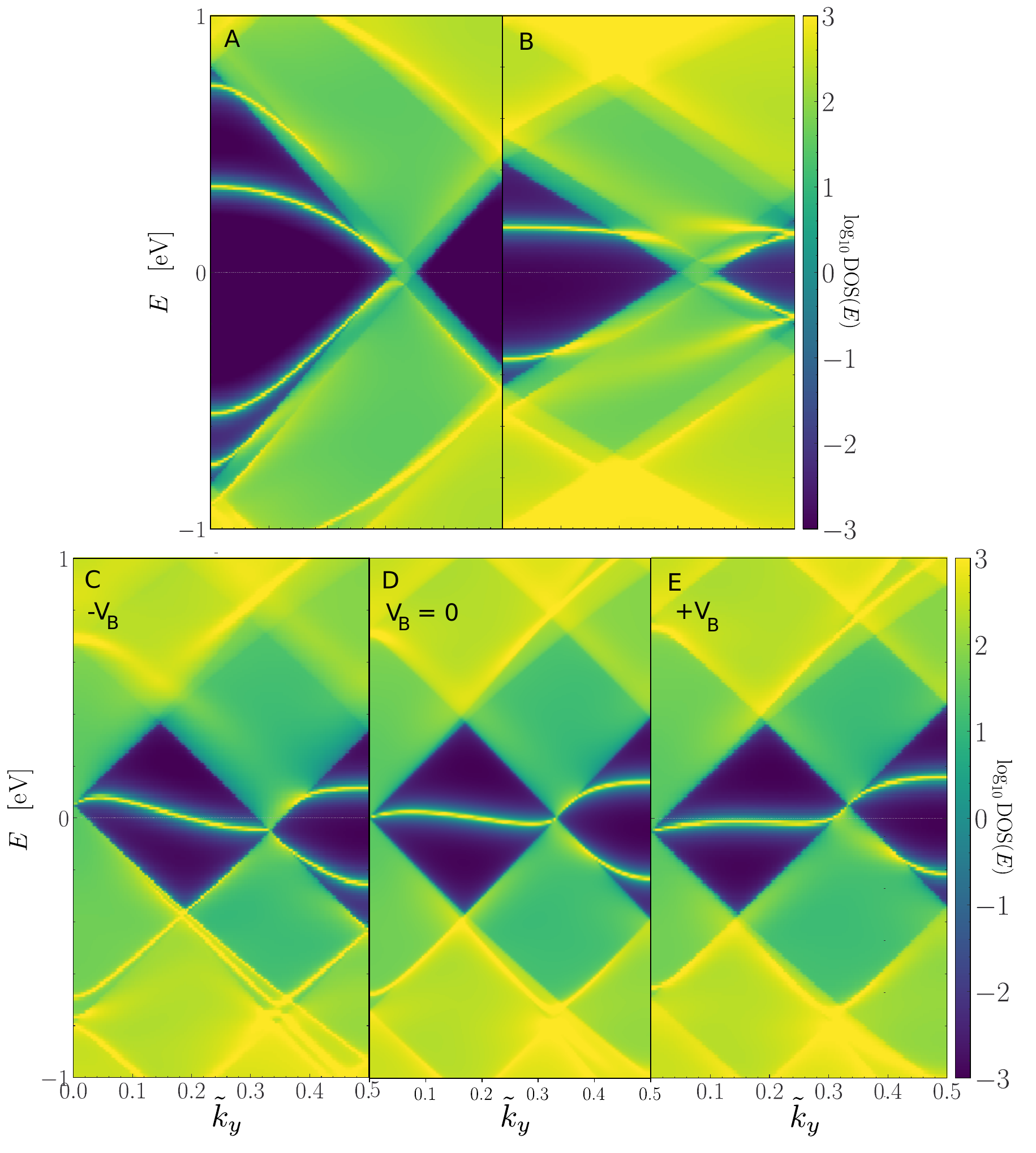}
    \caption{Non-equilibrium $k_y$-resolved DOS for a bias of $V_B=\mu_L-\mu_R=0.2$V shifting left/right chemical potentials. A) Nonequilibrium DOS of the GB from Fig. \ref{fig:4FinalStructures}A (Transparent). B) Nonequilibrium DOS of GB from Fig. \ref{fig:4FinalStructures}B (Opaque). C-E) The blocked (momentum mismatched) GB from Fig. \ref{fig:4FinalStructures}C at C) $V_B=-0.2$eV, D) at equilibrium $V_B=0$, and E) using a reversed bias $V_B=+0.2$eV. }
    \label{fig:NonEQ_DOS}
\end{figure}
However, in the transport gapped case, shown in Fig. \ref{fig:NonEQ_DOS}C-E, the LS bands are not seen to be broadened, but are rather modified and moving in position with bias. In particular, a bending of the LS band running from $\tilde{k}_y=0.0$ to $\tilde{k}_y=1/3$ appears as a consequence of the finite bias. This behavior is qualitatively different from Fig. \ref{fig:NonEQ_DOS}A and Fig. \ref{fig:NonEQ_DOS}B and stems from the different locations of the Dirac point in the TBZ. Thus, the bias over the GB changes the band-structure and group velocity of the transverse, localized one-dimensional GB state.
In a local transport measurement along the GB, this band-bending could manifest itself as an enhanced transversal conductivity when a bias is applied from the left to right. This in essence means an electric field in the $x$-direction will modify the local  conductivity close to the GB in the $y$-direction. 

The momentum mismatch between the electrodes surrounding the GB, leading to a gap in the transmission, may be lifted by phonon scattering supplying the missing momentum. This inelastic channel may, at finite bias, result in local Joule heating around the GB. This effect has been experimentally observed near GBs in graphene\cite{grosse2014direct}. We here qualitatively treat this phenomenon using the Special Thermal Displacement (STD) method\cite{gunst2017first}.
We use QuantumATK\cite{smidstrup2020,QATK} and the Brenner potential\cite{Brenner_2002} to generate the STD atomic displacements for various temperatures within a region of width $\pm 1$nm around the GB.
In order to include the right phonon momentum for a system with a gap corresponding to $\tilde{k}_y=1/3$, we consider phonons in a 3 times wider supercell. 

\begin{figure}
    \centering
    \includegraphics[width=0.8\linewidth]{ 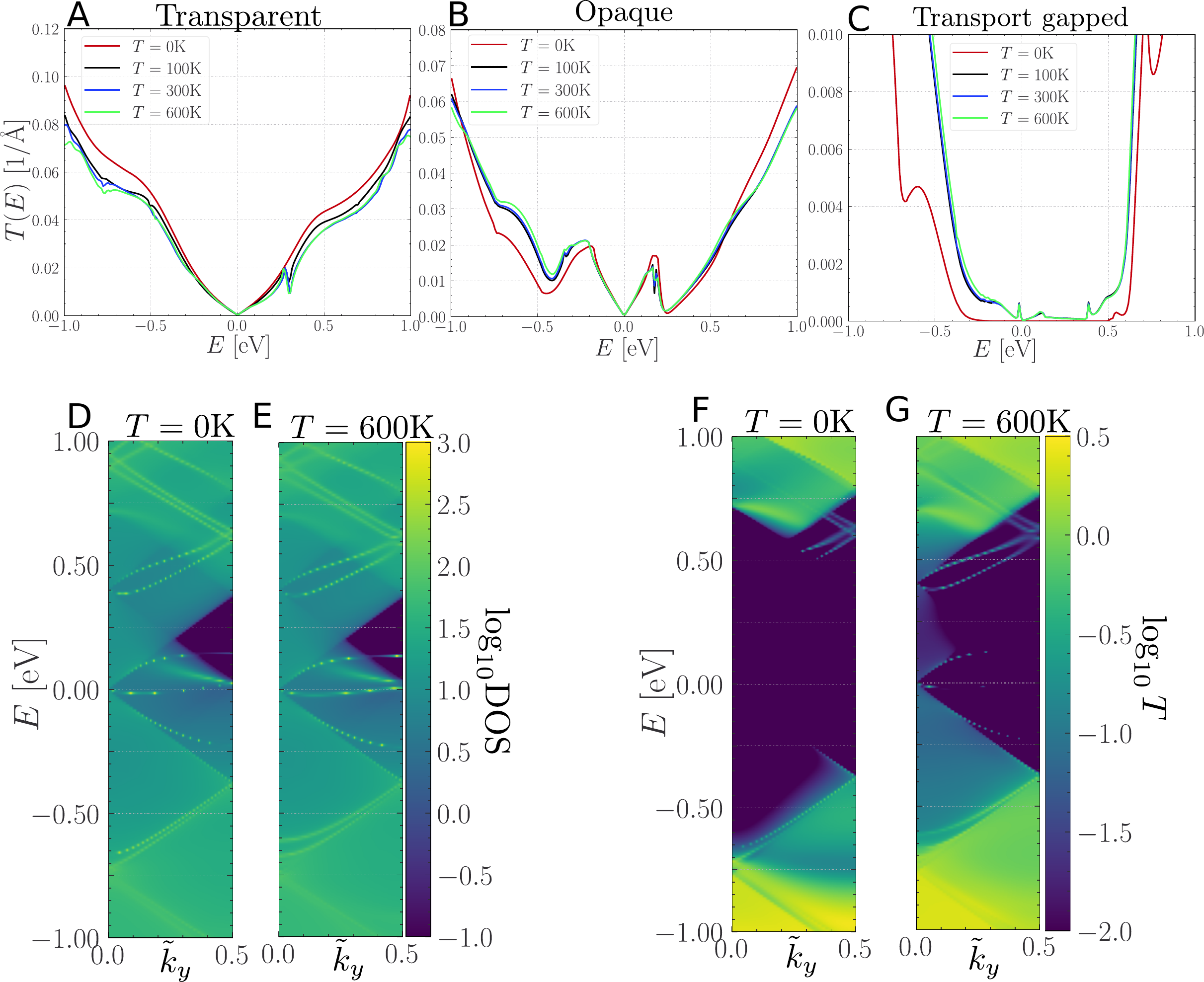}
    \caption{A,B,C) The $k_y$-integrated  transmission function with and without the STD at various temperatures. 
    D,E) The $k_y$-resolved DOS with and without the presence of a STD at $T=600$K using the structure from Fig. \ref{fig:4FinalStructures}C. F,G) The $k_y$-resolved transmission function with and without the presence of a STD at $T=600$K using the structure from Fig. \ref{fig:4FinalStructures}C. }
    \label{fig:Phonon}
\end{figure}

Using the STD method we can calculate the transmission function for the wider supercell, with and without the presence of phonons at a given temperature which, in turn, determines the statistical atomic displacements along the various phonons. We use this as a simple way to quantify the impact of the phonons on the conductive properties.  In Fig.~\ref{fig:Phonon}A,B,C the transmission function is shown at various temperatures with the STD. As the temperature rises, the pristine transmission function gets modified slightly by the presence of phonons. However, in the transport gapped case, the phonons furthermore have the effect of allowing current to run through the GB. This originates from the fact that the STD has a longer period than the primitive cell initially considered for the GB in Fig. \ref{fig:4FinalStructures}C, which allows mixing between the two Dirac cones from Fig. \ref{fig:2D_DOS}C that previously were momentum mismatched. The $E$ and $k_y$ resolved transmission function and DOS can furthermore be seen in Fig. \ref{fig:Phonon}D,E,F,G, and shows in detail how this longer periodicity induced by the STD results in a modified transmission function which has a visible imprint of the GB DOS.
The mechanism for altering the transmission function is the same also for the transparent and opaque cases, where the transmission function also gets modified when the DOS of the GB is large. For example, the dip from the STD in Fig. \ref{fig:Phonon}A around $E=0.3$eV is from the edge of the first conduction band shown in Fig. \ref{fig:2D_DOS}A.
%For this GB which within we are considering within the single-particle picture, does not conduct current at zero Kelvin (without the phonon) because of the momentum mismatch. However Fig. shows that when room-temperature is considered, there is a significant contribution to the transmission function from the phonon in what was before a transport gap. This is because the phonon breaks the translational symmetry and makes the period longer than $L_{GB}$. In Fig. \ref{fig:Phonon}A and \ref{fig:Phonon}B, the LS bands are furthermore seen to split up when phonon vibration is introduced.

%This in some ways would look like a nonzero $\sigma_{xy}$ but is coming from the fact that we are far from the linear regime where time-reversal symmetry dictates that this component should be zero.

\subsubsection{Effects of Electrostatic Gating}

\begin{figure}
  \centering
  \includegraphics[width=.85\linewidth]{ 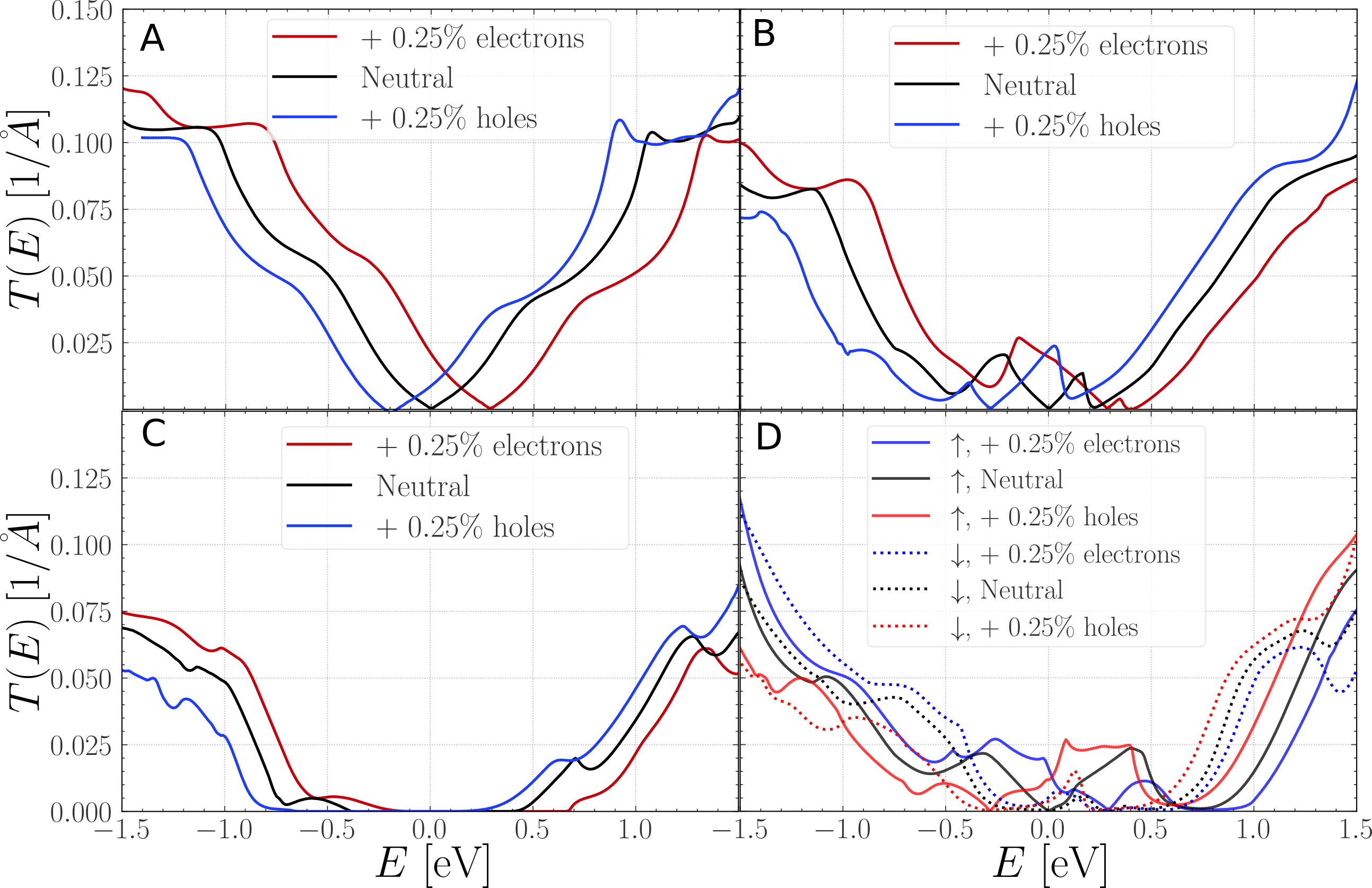}
  \caption{The transmission functions of the four example structures, calculated at different dopings indicated in the legend. Same labelling as in Fig. \ref{fig:4FinalStructures}. The percentage is relative to the number of carbon-nuclei. The doping percentage corresponds to a majority carrier concentration of $\approx 10^{13}$cm$^{-2}$   }
  \label{fig:Doping1}
\end{figure}

Graphene can be electrostatically gated using a back-gate (under or above the graphene sheet) to induce a change of charge in the graphene, shifting the Fermi level in the Dirac cone. Because the DOS of the bulk states in graphene will be the determining factor for where the Fermi-level will be located given a gate voltage, the Fermi-level in the GB will follow accordingly. We do not account for any re-relaxation of the GB from gating. Fig. \ref{fig:Doping1} shows the effects of electrostatic gating on transmission function the GBs in Fig. \ref{fig:4FinalStructures} where the Fermi-level is moving approximately $\pm0.3$eV from the charge neutrality point. 
The "transparent" GB from Fig. \ref{fig:4FinalStructures}A whose $k_y$-resolved DOS is seen in Fig. \ref{fig:2D_DOS}A is not subject to significant changes to its transmission function under introduction of additional holes or electrons by gating. The blocking GB in Fig. \ref{fig:Doping1}C displays only minor changes with gating. However, for both the "opaque" and "spin-polarized" GBs the transmission functions change significantly with gate voltage. The reason of the dips and peaks in the transmission function close to the charge neutrality point has previously in Fig. \ref{fig:2D_DOS} and Fig. \ref{fig:2D_Tr} been attributed to the presence of the localized states and their bands in the GB. 
\begin{figure}
  \centering
  \includegraphics[width=.8\linewidth]{ 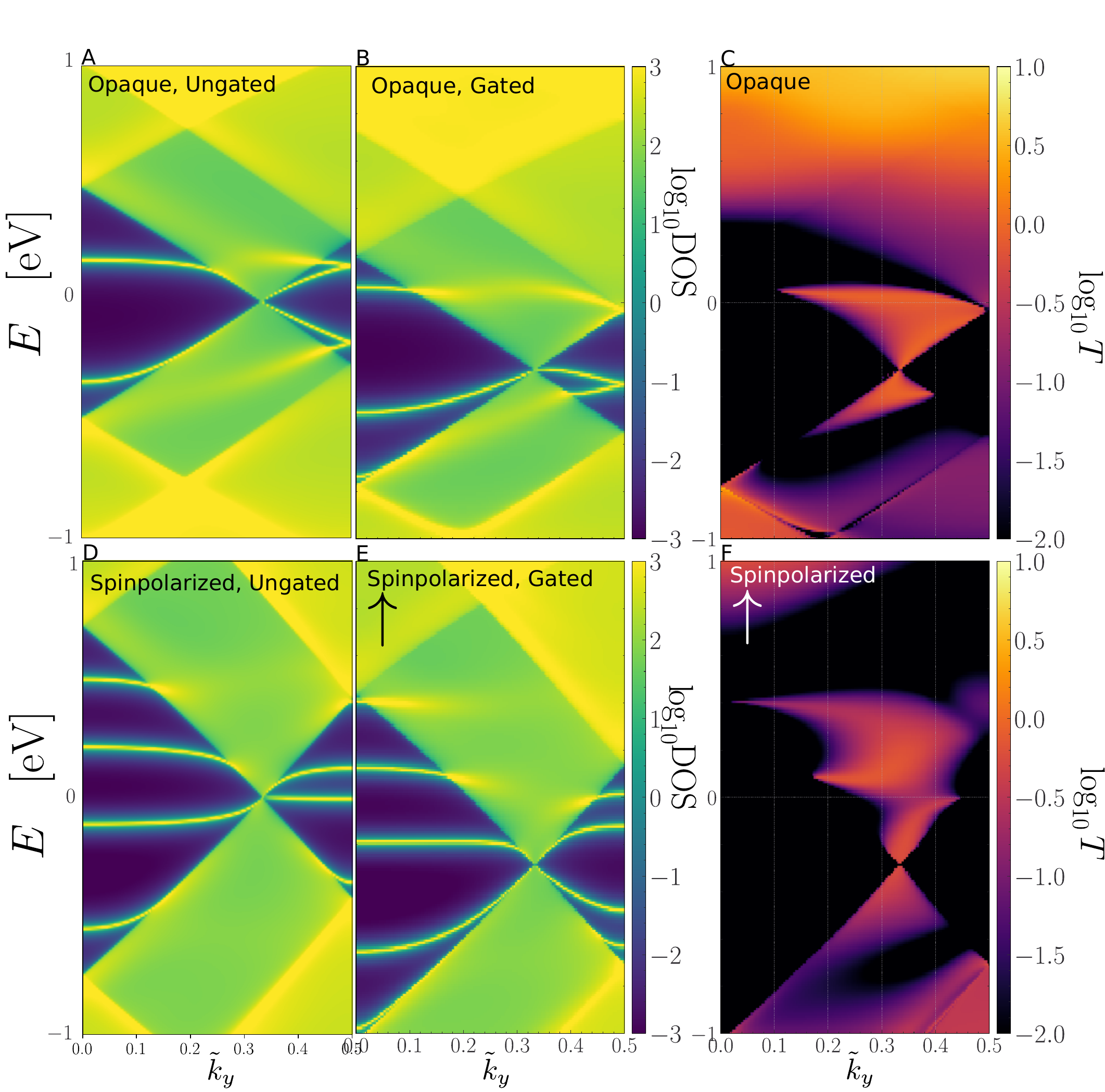}
  \caption{DOS (A+C) and transmission function  (B+D) for the "opaque" (A+B) and "spin-polarized" case(C+D) as a function of $k_y$, showing band-bending effects. Both cases gated such that there are $+0.25\%$ electrons per atom. A and D are repeated DOS from Fig. \ref{fig:2D_DOS}.}
  \label{fig:BandBending_1}
\end{figure}
When charge carriers are introduced into the GB, the bands bend significantly, as is evident from Fig. \ref{fig:BandBending_1}B and Fig. \ref{fig:BandBending_1}E. This band bending makes the bands intersect the Dirac cone in different places, giving rise to an altered transmission function as shown in Fig. \ref{fig:BandBending_1}C and Fig. \ref{fig:BandBending_1}F.
The results demonstrate how the transport for the large class of opaque GBs especially sensitive to gating.

\subsubsection{STM Simulation}

\begin{figure}
    \centering
    \includegraphics[width=0.7\linewidth]{ 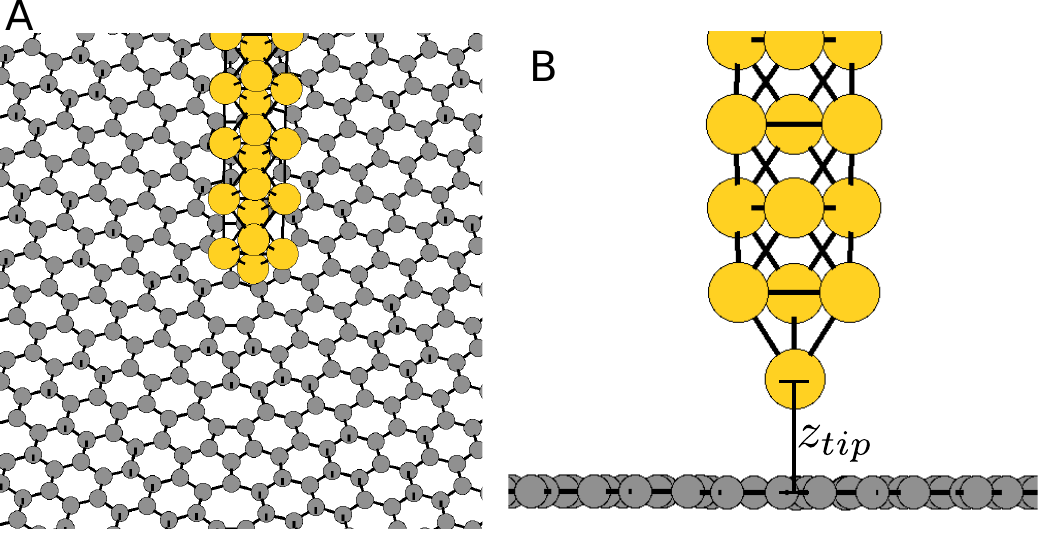}
    \caption{A) Top-view of the gold tip on graphene GB setup. B) side-view of the gold tip on graphene GB setup. The tip height $z_{tip}$ is seen. In this particular case it is the 5-7 GB structure from Fig. \ref{fig:4FinalStructures}A.} 
    \label{fig:STM_geom}
\end{figure}

\begin{figure}
    \centering
    \includegraphics[width=1.0\linewidth]{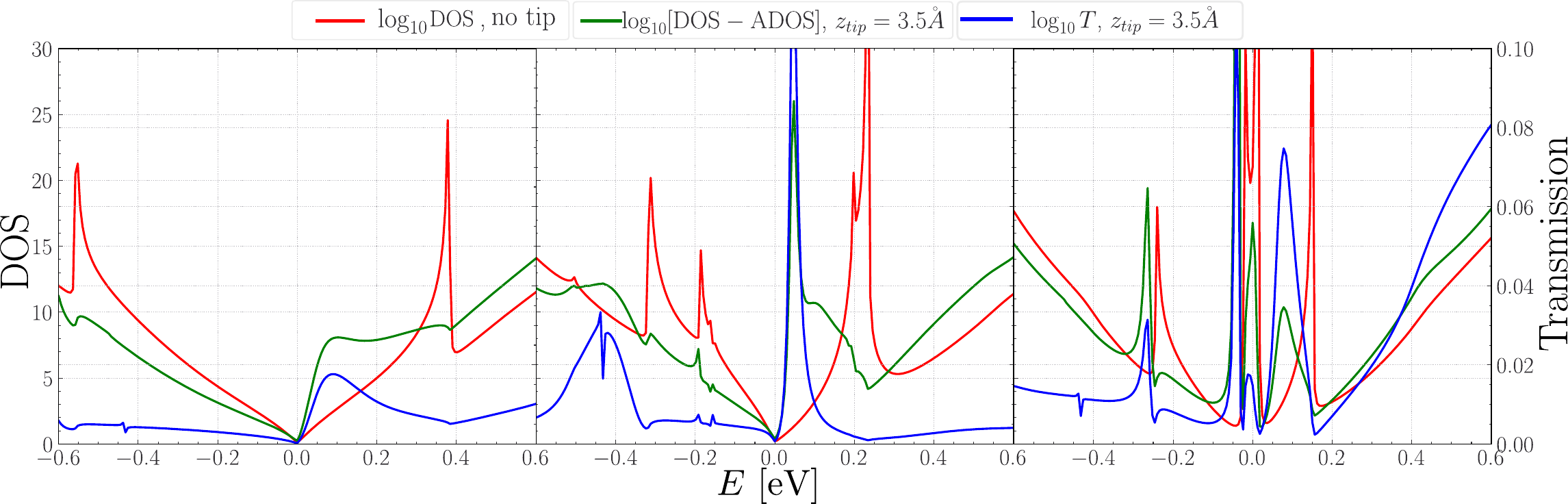}%{Tmp/stm_res_2.pdf}
    \caption{Plots of STM setup DOS minus gold electrode spectral DOS and transmission function from graphene to gold electrode. A) Results of GB in Fig. \ref{fig:4FinalStructures}A. B) Results of GB in Fig. \ref{fig:4FinalStructures}B. C) Results of GB in Fig. \ref{fig:4FinalStructures}C. ADOS is the spectral DOS of the gold electrode.}
    \label{fig:STM_sim}
\end{figure}

%As stated earlier, people have carried out STM measurements on GBs.\cite{luican2016localized} 
As mentioned above STM spectroscopy has been used to investigate the local density of states around GBs based on $\frac{\mathrm{d}I}{\mathrm{d}V}$-measurements\cite{luican2016localized, yang2014periodic, yin2014tuning,tison2014grain,ma2014evidence,li2020scanning}. The resulting $\frac{\mathrm{d}I}{\mathrm{d}V}$ curve is a measure of the local DOS right under the tip\cite{tersoff1985theory}.
Here we calculate the transmission function for a STM tip positioned a distance $z_{tip}$ over a GB. Here we use the real space self-energy from eq. \eqref{eq:RSSE} to model the infinite extent of the graphene sheet, and a regular gold tip\cite{li2021surface} to model a STM tip in contact with the graphene GB at a height of $z_{tip} = 3.5$\AA. 
This makes it possible to simulate a STM-measurement where there is a tunnel contact between the tip and the GB. The geometry is shown in Fig. \ref{fig:STM_geom} and for the DFT calculation a SZ basis is used. %and any spin-dependence in the gold-tip is neglected for ease of computation\MB{Does the Au-tip spin-polarize ? I would guess not.}.
When choosing how many times to tile the minimal cell for the Bloch folding of the Green's function in eq. \eqref{eq:RSSE}, the GB length is taken to be  $>3.5$ nm in all cases. The resulting transmission function, pristine GB DOS and GB DOS with tip is shown in Fig. \ref{fig:STM_sim}. 
The correspondence between the pristine DOS and peaks in the transmission functions is clear from these figures, even though the LS states broaden a bit when the tip is included in the calculation. Furthermore, the transmission function has sharp peaks in the unoccupied part of the spectrum for structures Fig. \ref{fig:4FinalStructures}A and Fig. \ref{fig:4FinalStructures}B indicating that the LS bands above the Fermi-level couple strongly to the gold tip. The states in the unoccupied part of the spectrum are however much less pronounced in the transmission function, even through we know from the DOS they are present. 

\begin{figure}
    \centering
    \includegraphics[width=0.75\linewidth]{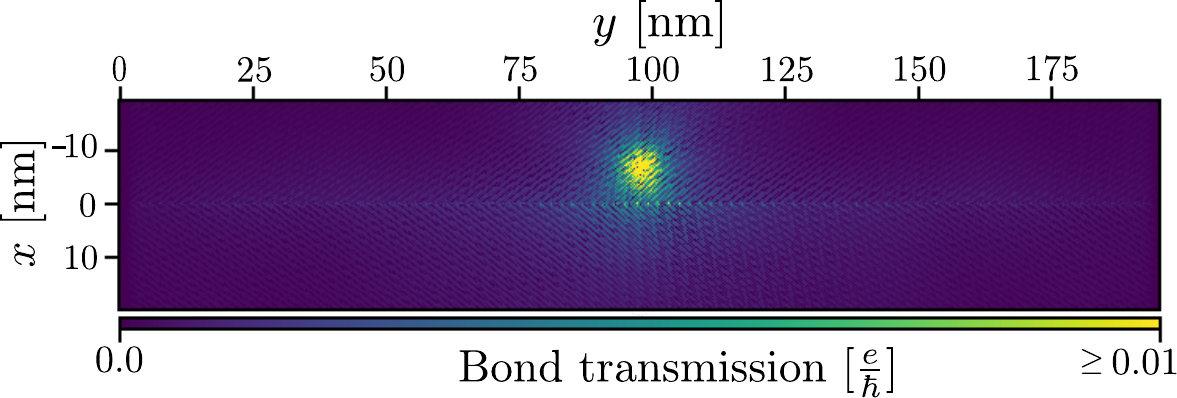}
    \caption{Bond-transmission for electron injection on a single site. GB is located at $x=0$. The GB structure is the same as in Fig. \ref{fig:4FinalStructures}A. The bond-transmissions are evaluated at $E = 0.05$eV.}
    \label{fig:BC}
\end{figure}

Large-scale tight-binding simulation (using parameters as in ref. \cite{calogero2018large}) is shown in Fig. \ref{fig:BC}. These demonstrate conduction along the GB for electrons injected on a single atomic site 7nm away from the GB (see SM. subsection \ref{sec:Largescale} for computational details).

\newpage
\subsubsection{Kubo-Greenwood approach to transport along the GB}

The conductivity of the GB in the $y$-direction related to the one-dimensional GB states is troublesome to evaluate by the Landauer formula, so instead a response function, $\sigma$, can be calculated using the Kubo-Greenwood formula, given as\cite{fan2021linear}
%\begin{align}
%    \label{eq:KuboCond}
%    \sigma_{y y}(E) =\frac{\hbar\mathrm{e}^2}{\pi A} \mathrm{Tr}[\mathbf{M}_{k_y}(E)\cdot %\mathbf{M}_{k_y}(E)] \quad \mathrm{with} \quad \mathbf{M}_{k_y}(E) = \frac{\mathbf{G}_{k_y}^{}(E) - [\mathbf{G}_{k_y}(E)] ^\dagger}{2}\mathbf{V}_{k_y}^y.
%\end{align}
\begin{align}
    \label{eq:KuboCond}
    \sigma_{y y}(E) %=\frac{\hbar\mathrm{e}^2}{\pi A} \mathrm{Tr}[\mathbf{M}(E)\cdot \mathbf{M}(E)]
    =\frac{G_0}{2\pi A_{cell}} \mathrm{Tr}\left[\langle\mathbf{L}_{k_y}(E)\cdot\mathbf{L}_{k_y}(E)\rangle_{k_y}\right],
\end{align}
with $A_{cell}$ being the unit cell area,  $\langle\phantom{.}\cdot\phantom{.}\rangle_{k_y}$ denoting the TBZ average, 
\begin{align}
\mathbf{L}_{k_y}(E) = \left[\frac{\mathbf{G}^r_{k_y}(E) - \mathbf{G}^a_{k_y}(E)}{2}\mathbf{\mathbf{v}}^y_{k_y}\right]
\end{align}
and $\mathbf{v}^y_{k_y}$ the transverse component of the velocity operator in the tight-binding representation, which can be taken as the ${k_y}$-derivative of the tight-binding Hamiltonian\cite{pedersen2001optical, fan2021linear}. $\sigma_{yy}$ describes the response to a static electric field as $J_y = \sigma_{yy}E_{y}$ where $J_y$ is the current running per unit cell of the simulation domain. The Green's function contains both bulk and GB contributions, where the bulk contribution will  increase with size of the cell chosen in the $x$-direction, but the GB part will not. 
This approach enables a quantification of the conductive properties of localized states with dispersion transverse to the standard transport direction (computational details are presented in SM. subsection \ref{sec:KuboCondCalc}).

\begin{figure}
    \centering
    \includegraphics[width=0.55\linewidth]{ 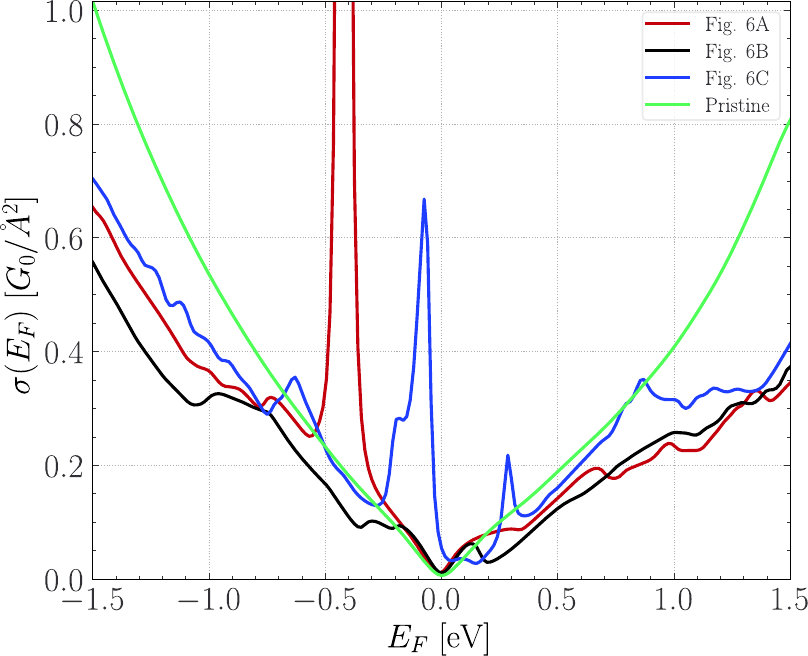}
    \caption{Transversal diffusive conductivity of three GBs compared to the conductivity of pristine graphene calculated in units of $G_0/A_{UC} = \frac{2\mathrm{e}^2}{hA_{UC}}$.}
    \label{fig:DiffCond}
\end{figure}

Using the Kubo-Greenwood formula eq. \eqref{eq:KuboCond}, a diffusive conductivity for the $y$-direction can be calculated for the same three GBs as in the STM simulations Fig. \ref{fig:STM_sim}. The Kubo-Greenwood formula in a periodic solid evaluates to a sum over $k$-derivatives of the available bands at the considered energy\cite{taghizadeh2017linear}. In the present case, there are additional LS band(s) available, which lead to notable peaks in the transverse conductivity in Fig. \ref{fig:DiffCond}. Here the peaks are instead located in the occupied part of the spectrum. It is also noteworthy that the GB with the transport gap a nonzero conductivity at $E=0$ coming from the flat-band in Fig. \ref{fig:2D_DOS}C.

\section{Discussion}
The bands that lie outside the graphene Dirac cones are a recurring feature of all the GBs and the different types of band-structures from Fig. \ref{fig:2D_DOS} should therefore be possible to observe experimentally. The LS bands may also show up as polarization dependence in optical measurements using scanning probes, since the electric field below the tip can point along the GB or perpendicular to it. In the first case the localized states can conduct current along it, while being just a scatterer in the second case. This type of experiment has already been carried out in ref. \cite{fei2013electronic} and might also show additional features related to polarization direction and gating as well as depend on the class of GB, c.f. Fig. \ref{fig:2D_DOS} and Fig. \ref{fig:BandBending_1}.

The phenomenon of the GBs conducting current well along the GB shown in Fig. \ref{fig:DiffCond} could also be detectable by scanning magnetometry to map the local current density as has been done in e.g. ref. \cite{tetienne2017quantum}. This type of experiment should show a large current density at the GB if the Fermi-level can be adjusted to match one of the peaks in Fig. \ref{fig:DiffCond}.

%\textcolor{blue}{\textit{17/11/23}:
Furthermore, as described in refs. \cite{srivastava2018trade, wood2014first}, quantum capacitance measurements are directly related to the electronic density of states. This in turn means the LS bands might be detectable in quantum capacitance measurements of samples which are densely populated by a certain type of GBs. In particular the transport gapped GB from Fig. \ref{fig:2D_DOS}C and the band running from one Dirac cone to the other might show a significant peak in its quantum capacitance at $V=0$, given that there are enough of these GBs to contribute significantly to the total sample DOS. This has also been calculated in relation to GBs in ref. \cite{jimenez2014impact}. \\
Evidence of Peierls instability\cite{peierls1955quantum, heeger1988solitons} has also been found in TMD GBs \cite{batzill2018mirror} but there is (the the authors' knowledge) very limited evidence in graphene GBs\cite{capasso2014graphene}. It is however a possibility because a periodic lattice distortion could in some cases lower the total energy. We have however not made any considerations of longer-range relaxation when creating the workflow from Fig. \ref{fig:Flow} and relaxing the GBs.
%}

The band-bending induced by gating that was predicted in Fig. \ref{fig:BandBending_1} should also be experimentally detectable with an STM measurement, since the van Hove singularities of the band edges have already been detected and during gating these peaks should move.  The fact that the band structure of the GB states in some cases can be manipulated could make GBs a test-bed for exotic physics, such as a flat band that is tuneable by bias, see Fig. \ref{fig:NonEQ_DOS}C and Fig. \ref{fig:NonEQ_DOS}D or a gate tunable spin-filter, see Fig. \ref{fig:J_scatter}B and Fig. \ref{fig:Doping1}.

The structures generated in this work are created without the inclusion of contaminants such as water and other specimens present during the CVD growth. Molecules can however subsequently adsorb onto the graphene surface during storage, device processing or measurement\cite{brito2012b,seifert2015role}. In some particular cases, such as hydrogen and some other species, it is favorable to make bonds to carbon atoms inside the GB\cite{brito2011hydrogenated,zhang2014grain}. The inclusion of such adsorbates, while interesting, would many-double the number of calculations needed to quantify the effects as there are different outcomes depending on where the absorbate bonds in the GB\cite{vargas2017grain}. However, identifying which GBs belong to which of the four classes listed in Table 2 is useful for knowing which GBs to select for a given property or function of a graphene device, even through it may be modified . Followingly growth optimizing conditions for generating a higher prevalence of this type of GB could then be carried out.

Another unknown factor that could impact validity of the modeling in this work is the fact that the GB phonons couples to some degree to the electrons around the GB as was demonstrated in Fig. \ref{fig:Phonon}. It is therefore also plausible that inelastic effects could significantly modify the ability of the GBs to conduct current. However doing the inelastic calculations have been outside of computational feasibility for the structures considered in this paper, but as previously stated, Joule-heating have been observed at GBs\cite{grosse2014direct}, which is an inelastic phenomenon. 

\section{Conclusions}
An algorithm for automated GB generation has been constructed and used to build a dataset containing \TotalStructs GB structures (Fig. \ref{fig:1}). These GBs have been grouped into four categories according to their ability to conduct (spin-polarized) current (Fig. \ref{fig:J_scatter} and Table 2). The conductive properties are shown to be linked to the local electronic structure at the GB, namely the LS states existing there, which in some cases are a major inhibitor of the GBs ability to conduct current. These LS states can in some cases be significantly affected by charge doping, and in the case of a transport-gapped GB, be manipulated by applying a bias to the device. The conductive properties have been characterised in several ways, using both a regular transport setup for conduction directly across the GB (Fig. \ref{fig:2D_Tr}) in conjunction with electrostatic gating (Fig. \ref{fig:Doping1} and Fig. \ref{fig:BandBending_1}), in addition to with STM-simulations (Fig. \ref{fig:STM_sim}) and diffusive conductivity calculations (Fig. \ref{fig:DiffCond}). The electronic structure has furthermore been characterised in equilibrium (Fig. \ref{fig:2D_DOS}, Fig. \ref{fig:QPstate} and Fig. \ref{fig:BandBending_1}) and out-of-equilibrium (Fig. \ref{fig:NonEQ_DOS}). 

This work is our attempt at providing a detailed classification of graphene GBs. This should be useful for understanding and reducing detrimental effects and for creating unique transport properties in nanoscale systems with potential for device applications.
%\section{Acknowledgements}
%The authors acknowledge ...
\section{Data availability Statement}
The structures used for this article and code used to generate these can be found in ref. \cite{githubGB} and further data can be retrieved by request at abalo@dtu.dk by email. 
\section{Conflict of interest}
The authors declare no competing financial or non-financial interests. 
\section{Funding statement}
This research was funded by Independent Research Fund Denmark, grant number
0135-00372A.
\printbibliography
\newpage
\section{Supplementary Materials}

\subsection{Cost function for Geometry Creation}\label{sec:Costfunction}
A purely numerical cost function on the form of a Morse potential modified with a nearest neighbor angular dependence is used:
\begin{align}
V(\{\mathbf{r}_i \})&= \sum_{<i,j>} D_e[(1 - \mathrm{e}^{a(d_{ij} - d_{CC})})^2 - 1]  \\&+ \sum_{i } \sum_{m,n \in \mathrm{NN\phantom{.}of\phantom{.}i}|m>n}A_w f_D(|\mathbf{r}_{in}|+ |\mathbf{r}_{im}| - A_f d_{CC},s)[(\frac{|(\arccos(\frac{\mathbf{r}_{im}\cdot\mathbf{r}_{in}}{|\mathbf{r}_{in}| |\mathbf{r}_{im}|}) - \frac{2\pi}{3})|}{A_0})^p - \theta_{0}]
\end{align}
with $dij = |\Vec{r}_i - \Vec{r}_j|$, $D_e=4, A_w = 1.0, A_f = 2.3, A_0 = 16\pi/180, p=2, \theta_0 = 0$, $a = 2$, $d_{CC}=1.4202\textup{~\AA}$, $s=0.08$  and 12 nearest neighbors used. Distances are measured in $\textup{~\textup{~\AA}}$. This cost function is used to determine whenever to remove or add a carbon atom at various places, and has been constructed empirically for the purpose of yielding reasonable initial guesses for the GB structures. The condition for placing an atom is if $V(\{\mathbf{r}_i \} \cup \{\mathrm{r}_{new}\})<V(\{\mathbf{r}_i \})$. The first part of the potential takes care of the distance dependence of the energy when the distance between carbon atoms are changed, and the second part takes makes the result tend to form bonds are closer to $120^\circ$ as is the case in the hexagonal lattice of graphene. 

\subsection{Computational Details for Real Space Self Energy}\label{sec:RSSEcalc}
The $k_y$-integral in eq. \eqref{eq:RSSE} is calculated using the adaptive integration routine quad\_vec of the scipy integrate module, NumPy arrays and the Numba package\cite{2020SciPy-NMeth, harris2020array, lam2015numba}. An upfolding of the minimal cell greens function can furthermore be employed in this calculation\cite{papior2019removing}. This calculation has been automated in the siesta\_python code (ref. \cite{siesta_python_web}) in an energy-parallel fashion and with the possibility for left and right electrodes that are different in number of orbitals and possibly electronic structure. The siesta\_python code mainly consists of an ASE-inspired\cite{larsen2017atomic, gjerding2021atomic} python class for handling the writing, processing and reading input-files for electronic structure and transport codes, primarily the SIESTA and TBtrans codes.  

\subsection{Computational Details for Diffusive Transverse Conductivity}\label{sec:KuboCondCalc}
The Greens function going into eq. \eqref{eq:KuboCond} can furthermore be calculated using the efficient algorithm for calculating the inverse of a block-tridiagonal matrix\cite{papior2017improvements,reuter2012efficient}. This algorithm is available in the Block\_matrices code (See ref. \cite{Block_matrices_web}). An optimized pivoting scheme for making the matrix block-tridiagonal is obtained using the reverse Cuthill-McKee algorithm and the TBZ average is calculated using an adaptive integrator, both from SciPy\cite{cuthill1969reducing, 2020SciPy-NMeth}. Wrapper functions are available in the siesta\_python code (ref. \cite{siesta_python_web}).

%\subsection{Naive Transmission Function}
%A more naive approach compared to the Landauer formula is to just consider the available number of scattering states $\kappa_\alpha(E)$ in the two electrodes at a particular energy $E$ and define the naive transmission as
%\begin{align}
%    \label{eq:naivetransmission}
%    K^{i\rightarrow j}(E) = \min_{\alpha} \{\kappa_\alpha(E) \}
%\end{align}
%which neglects the any of the before-mentioned phenomena the Landauer formula accounts for. However by comparing $K^{i \rightarrow j }$ to the Landauer formula, the influence of the  atomic structure on the electrons moving through the device can be assessed. $K(E)$ is also the maximal value the Landauer transmission $T(E)$ can take. 
\subsection{Large-scale Simulations}\label{sec:Largescale}
For the bond-current calculations in Fig. \ref{fig:BC} a simple tight-binding model using a nearest-neighbor model with hopping $t=-2.7$eV with CAPs having been placed at $y<d_{CAP}$ and  $y>L_y - d_{CAP}$\cite{calogero2018large}.
The tip self-energy is modelled as a single site chain coupling only to one carbon site. The hopping element is $t_{chain}= -2.7$eV making it effectively a wide-band self-energy in the energy range considered. 

\end{document}